\documentclass[journal]{IEEEtran}
\ifCLASSINFOpdf
\else
\fi
\usepackage{textcomp} 
\usepackage{adjustbox} 
\usepackage{textcase}
\usepackage{amsfonts}
\usepackage{graphicx}

\RequirePackage{xspace}
\RequirePackage{graphics}
\usepackage{xcolor}
\RequirePackage{textcomp}
\usepackage{mathrsfs}
\usepackage{lipsum}
\usepackage{listings}
\usepackage{multirow}
\usepackage{array}
\usepackage{pifont} 
\usepackage{stmaryrd}
\usepackage{footnote}
\usepackage{makecell}
\usepackage{xparse}

\usepackage{cite}

\usepackage{amssymb}
\usepackage{graphicx}
\usepackage{amsmath}
\usepackage{amsfonts}
\RequirePackage{xspace}
\RequirePackage{graphics}
\usepackage{xcolor}
\RequirePackage{textcomp}
\usepackage{keyval}
\usepackage{xspace}
\usepackage{mathrsfs}
\usepackage{paralist}
\usepackage{lipsum}
\usepackage{listings}
\usepackage{multirow}
\usepackage{array}
\usepackage{pifont} 
\usepackage{stmaryrd}
\usepackage{adjustbox} 
\usepackage{textcase}
\usepackage{makecell}
\usepackage{flushend} 

\RequirePackage{tikz}
\usetikzlibrary{arrows,automata,shapes,calc,through,decorations.pathmorphing,decorations.fractals,chains,shapes.multipart}

\usepackage{arydshln} 
\usepackage[free-standing-units]{siunitx}

\usepackage{circuitikz}

\def\titlename{Virtual Prototyping and Distributed Control for Solar Array with Distributed Multilevel Inverter\xspace}
\def\authorluan{Luan Viet Nguyen}
\def\authortaylor{Taylor T Johnson}

\newcommand{\nnnum}[1]{\relax\ifmmode 
  {\mathbb #1}_{\geq 0} \else ${\mathbb #1}_{\geq 0}$
  \fi}
\newcommand{\npnum}[1]{\relax\ifmmode 
  {\mathbb #1}_{\leq 0} \else ${\mathbb #1}_{\leq 0}$
  \fi}
\newcommand{\pnum}[1]{\relax\ifmmode 
  {\mathbb #1}_{> 0} \else ${\mathbb #1}_{> 0}$
  \fi}
\newcommand{\nnum}[1]{\relax\ifmmode 
  {\mathbb #1}_{< 0} \else ${\mathbb #1}_{< 0}$
  \fi}
\newcommand{\plnum}[1]{\relax\ifmmode 
  {\mathbb #1}_{+} \else ${\mathbb #1}_{+}$
  \fi}
\newcommand{\nenum}[1]{\relax\ifmmode 
  {\mathbb #1}_{-} \else ${\mathbb #1}_{-}$
  \fi}

\newcommand{\extb}[1]{\relax\ifmmode {\sf ExtBeh}_{#1} \else ${\sf ExtBeh}_{#1}$\fi} 
\newcommand{\tdists}[1]{\relax\ifmmode {\sf Tdists}_{#1} \else ${\sf Tdists}_{#1}$\fi} 

\newcommand{\exec}[1]{\relax\ifmmode {\sf Execs}_{#1} \else ${\sf Exec}_{#1}$\fi} 
\newcommand{\execf}[1]{\relax\ifmmode {\sf Execs}^*_{#1} \else ${\sf Exec}^*_{#1}$\fi} 
\newcommand{\execi}[1]{\relax\ifmmode {\sf Execs}^\omega_{#1} \else ${\sf Exec}^\omega_{#1}$\fi} 

\newcommand{\ctrace}[1]{\relax\ifmmode {\sf Ctraces}_{#1} \else ${\sf Ctraces}_{#1}$\fi} 

\newcommand{\trace}[1]{\relax\ifmmode {\sf Traces}_{#1} \else ${\sf Traces}_{#1}$\fi} 
\newcommand{\tracef}[1]{\relax\ifmmode {\sf Traces}^*_{#1} \else ${\sf Traces}^*_{#1}$\fi} 
\newcommand{\tracei}[1]{\relax\ifmmode {\sf Traces}^\omega_{#1} \else ${\sf Traces}^\omega_{#1}$\fi} 

\newcommand{\frag}[1]{\relax\ifmmode {\sf Frags}_{#1} \else ${\sf Frags}_{#1}$\fi} 
\newcommand{\fragf}[1]{\relax\ifmmode {\sf Frags}^*_{#1} \else ${\sf Frags}^*_{#1}$\fi} 
\newcommand{\fragi}[1]{\relax\ifmmode {\sf Frags}^\omega_{#1} \else ${\sf Frags}^\omega_{#1}$\fi} 

\newcommand{\reach}[1]{\relax\ifmmode {\sf Reach}_{#1} \else ${\sf Reach}_{#1}$\fi}

\newcommand{\type}[1]{\ms{type{(#1)}}}

\newcommand{\restrrange}{\mathrel{\downarrow}}

\def\A{{\cal A}} 
\def\D{{\cal D}} 
\def\E{{\cal E}} 
\def\I{{\cal I}} 
\def\R{{\cal R}} 
\def\T{{\cal T}} 
\def\U{{\cal U}} 

\newcommand{\col}[1]{\relax\ifmmode \mathscr #1\else $\mathscr #1$\fi}

\definecolor{HIOAcolor}{rgb}{0.776,0.22,0.07}

\newcommand{\SC}[2]{\relax\ifmmode {\tt Scount}(#1,#2) \else ${\tt Scount}(#1,#2)$\fi} 
\newcommand{\SCM}[2]{\relax\ifmmode {\tt Smin}(#1,#2) \else ${\tt Smin}(#1,#2)$\fi} 
\newcommand{\Aut}[1]{\relax\ifmmode {\tt Aut}(#1) \else ${\tt Aut}(#1)$\fi}

\newcommand{\act}[1]{{\operatorname{\mathsf{#1}}}}

\newcommand{\deq}{\mathrel{\stackrel{\scriptscriptstyle\Delta}{=}}}

\newcommand{\seclabel}[1]{\label{sec:#1}}
\newcommand{\secref}[1]{Section~\ref{sec:#1}}

\newcommand{\figlabel}[1]{\label{fig:#1}}
\newcommand{\figref}[1]{Figure~\ref{fig:#1}}
\newcommand{\figreftwo}[2]{Figures~\ref{fig:#1}~and~\ref{fig:#2}}

\newcommand{\figreffour}[4]{Figures~\ref{fig:#1},~\ref{fig:#2},~\ref{fig:#3},~and~\ref{fig:#4}}
\newcommand{\tablabel}[1]{\label{tab:#1}}
\newcommand{\tabref}[1]{Table~\ref{tab:#1}}

\newcommand{\eqlabel}[1]{\label{eq:#1}}
\renewcommand{\eqref}[1]{Equation~\ref{eq:#1}}

\newcommand{\remove}[1]{}
\newcommand{\salg}[1]{\relax\ifmmode {\mathcal F}_{#1}\else ${\mathcal F}_{#1}$\fi} 
\newcommand{\msp}[1]{\relax\ifmmode (#1, \salg{#1}) \else $(#1, \salg{#1})$\fi} 
\newcommand{\msprod}[2]{\relax\ifmmode ( #1 \times #2, \salg{#1} \otimes \salg{#2}) \else $(#1 \times #2, \salg{#1} \otimes \salg{#2})$\fi} 
\newcommand{\dist}[1]{\relax\ifmmode {\mathcal P}\msp{#1}
  \else ${\mathcal P}\msp{#1}$\fi} 
\newcommand{\subdist}[1]{\relax\ifmmode {\mathcal S}{\mathcal P}\msp{#1} 
  \else ${\mathcal S}{\mathcal P}\msp{#1}$\fi} 
\newcommand{\disc}[1]{\relax\ifmmode {\sf Disc}(#1)
  \else ${\sf Disc}(#1)$\fi} 

\newcommand{\Trajeq}{\relax\ifmmode {\mathcal R}_\T \else ${\mathcal R}_\T$\fi} 
\newcommand{\Acteq}{\relax\ifmmode {\mathcal R}_A \else ${\mathcal R}_A$\fi} 
\newcommand{\noop}{\relax\ifmmode \lambda \else $\lambda$\fi} 
\newcommand{\close}[1]{\relax\ifmmode \overline{#1} \else $\overline{#1}$\fi}

\newcommand{\fstate}{\mathop{\mathsf {fstate}}}
\newcommand{\lstate}{\mathop{\mathsf {lstate}}}
\newcommand{\ltime}{\mathop{\mathsf {ltime}}}

\newcommand{\ie}{i.e.,\xspace}

\newcommand{\eg}{e.g.,\xspace}

\newcommand{\tup}[1]
           {
             \relax\ifmmode
             \langle #1 \rangle
             \else $\langle$ #1 $\rangle$ \fi
           }

\newcommand{\lit}[1]{ \relax\ifmmode
                \mathord{\mathcode`\-="702D\sf #1\mathcode`\-="2200}
                \else {\it #1} \fi }

\newcommand{\figuresize}{\scriptsize}

\lstdefinelanguage{ioa}{
  basicstyle=\figuresize,
  keywordstyle=\bf \figuresize,
  identifierstyle=\it \figuresize,
  emphstyle=\tt \figuresize,
  mathescape=true,
  tabsize=20,
  sensitive=false,
  columns=fullflexible,
  keepspaces=false,
  flexiblecolumns=true,
  basewidth=0.05em,
  escapeinside={(*@}{@*)},
  moredelim=[il][\rm]{//},
  moredelim=[is][\sf \figuresize]{!}{!},
  moredelim=[is][\bf \figuresize]{*}{*},
  keywords={automaton,and, 
  	 choose,const,continue, components,
  	 discrete, do,
  	 eff, Eff, external,else, elseif, evolve, end,
  	 fi,for, forward, from,
  	 hidden,
  	 in,input,internal,if,invariant, initially, imports,
     let,
     or, output, operators, od, of,
     pre, Pre,
     return,
     such,satisfies, stop, signature, simulation, 
     trajectories,trajdef, transitions, that,then, type, types, to, tasks,
     variables, vocabulary, 
     when,where, with,while},
  emph={set, seq, tuple, map, array, enumeration},   
   literate=
        {(}{{$($}}1
        {)}{{$)$}}1
        {\\in}{{$\in\ $}}1
        {\\preceq}{{$\preceq\ $}}1
        {\\subset}{{$\subset\ $}}1
        {\\subseteq}{{$\subseteq\ $}}1
        {\\supset}{{$\supset\ $}}1
        {\\supseteq}{{$\supseteq\ $}}1
        {\\forall}{{$\forall$}}1
        {\\le}{{$\le\ $}}1
        {\\ge}{{$\ge\ $}}1
        {\\gets}{{$\gets\ $}}1
        {\\cup}{{$\cup\ $}}1
        {\\cap}{{$\cap\ $}}1
        {\\langle}{{$\langle$}}1
        {\\rangle}{{$\rangle$}}1
        {\\exists}{{$\exists\ $}}1
        {\\bot}{{$\bot$}}1
        {\\rip}{{$\rip$}}1
        {\\emptyset}{{$\emptyset$}}1
        {\\notin}{{$\notin\ $}}1
        {\\not\\exists}{{$\not\exists\ $}}1
        {\\ne}{{$\ne\ $}}1
        {\\to}{{$\to\ $}}1
        {\\implies}{{$\implies\ $}}1
        {<}{{$<\ $}}1
        {>}{{$>\ $}}1
        {=}{{$=\ $}}1
        {~}{{$\neg\ $}}1
        {|}{{$\mid$}}1
        {'}{{$^\prime$}}1
        {\\A}{{$\forall\ $}}1
        {\\E}{{$\exists\ $}}1
        {\\nE}{{$\nexists\ $}}1
        {\\/}{{$\vee\,$}}1
        {\\vee}{{$\vee\,$}}1
        {/\\}{{$\wedge\,$}}1
        {\\wedge}{{$\wedge\,$}}1
        {=>}{{$\Rightarrow\ $}}1
        {->}{{$\rightarrow\ $}}1
        {<=}{{$\Leftarrow\ $}}1
        {<-}{{$\leftarrow\ $}}1
        {~=}{{$\neq\ $}}1
        {\\U}{{$\cup\ $}}1
        {\\I}{{$\cap\ $}}1
        {|-}{{$\vdash\ $}}1
        {-|}{{$\dashv\ $}}1
        {<<}{{$\ll\ $}}2
        {>>}{{$\gg\ $}}2
        {||}{{$\|$}}1
        {[}{{$[$}}1
        {]}{{$\,]$}}1
        {[[}{{$\langle$}}1
        {]]]}{{$]\rangle$}}1
        {]]}{{$\rangle$}}1
        {<=>}{{$\Leftrightarrow\ $}}2
        {<->}{{$\leftrightarrow\ $}}2
        {(+)}{{$\oplus\ $}}1
        {(-)}{{$\ominus\ $}}1
        {_i}{{$_{i}$}}1
        {_j}{{$_{j}$}}1
        {_{i,j}}{{$_{i,j}$}}3
        {_{j,i}}{{$_{j,i}$}}3
        {_0}{{$_0$}}1
        {_1}{{$_1$}}1
        {_2}{{$_2$}}1
        {_n}{{$_n$}}1
        {_p}{{$_p$}}1
        {_k}{{$_n$}}1
        {-}{{$\ms{-}$}}1
        {@}{{}}0
        {\\delta}{{$\delta$}}1
        {\\R}{{$\R$}}1
        {\\Rplus}{{$\Rplus$}}1
        {\\N}{{$\N$}}1
        {\\times}{{$\times\ $}}1
        {\\tau}{{$\tau$}}1
        {\\alpha}{{$\alpha$}}1
        {\\beta}{{$\beta$}}1
        {\\gamma}{{$\gamma$}}1
        {\\ell}{{$\ell\ $}}1
        {--}{{$-\ $}}1
        {\\TT}{{\hspace{1.5em}}}3        
      }

\lstdefinelanguage{ioaNums}[]{ioa}
{
  numbers=left,
  numberstyle=\tiny,
  stepnumber=2,
  numbersep=4pt
}

\lstdefinelanguage{ioaNumsRight}[]{ioa}
{
  numbers=right,
  numberstyle=\tiny,
  stepnumber=2,
  numbersep=4pt
}

\lstnewenvironment{IOA}%
  {\lstset{language=IOA}}
  {}

\lstnewenvironment{IOANums}%
  {
  \if@firstcolumn
    \lstset{language=IOA, numbers=left, firstnumber=auto}
  \else
    \lstset{language=IOA, numbers=right, firstnumber=auto}
  \fi
  }
  {}

\lstnewenvironment{IOANumsRight}%
  {
    \lstset{language=IOA, numbers=right, firstnumber=auto}
  }
  {}

\newcommand{\linefigioa}[9]{

}

\lstdefinelanguage{ioaLang}{%
  basicstyle=\ttfamily\small,
  keywordstyle=\rmfamily\bfseries\small,
  identifierstyle=\small,
  keywords={assumes,automaton,axioms,backward,bounds,by,case,choose,components,const,d,det,discrete,do,eff,else,elseif,ensuring,enumeration,evolve,fi,fire,follow,for,forward,from,hidden,if,in,%
    input,initially,internal,invariant,let, local,od,of,output,pre,schedule,signature,so,%
    simulation,states,variables, tasks, stop,tasks,that,then,to,trajdef,trajectory,trajectories,transitions,tuple,type,union,urgent,uses,when,where,while,yield},
  literate=
        {\\in}{{$\in$}}1
        {\\preceq}{{$\preceq$}}1
        {\\subset}{{$\subset$}}1
        {\\subseteq}{{$\subseteq$}}1
        {\\supset}{{$\supset$}}1
        {\\supseteq}{{$\supseteq$}}1
        {\\rho}{{$\rho$}}1
        {\\infty}{{$\infty$}}1
        {<}{{$<$}}1
        {>}{{$>$}}1
        {=}{{$=$}}1
        {~}{{$\neg$}}1 
        {|}{{$\mid$}}1
        {'}{{$^\prime$}}1
        {\\A}{{$\forall$}}1 {\\E}{{$\exists$}}1
        {\\/}{{$\vee$}}1 {/\\}{{$\wedge$}}1 
        {=>}{{$\Rightarrow$}}1 
        {->}{{$\rightarrow$}}1 
        {<=}{{$\leq$}}1 {>=}{{$\geq$}}1 {~=}{{$\neq$}}1
        {\\U}{{$\cup$}}1 {\\I}{{$\cap$}}1
        {|-}{{$\vdash$}}1 {-|}{{$\dashv$}}1
        {<<}{{$\ll$}}2 {>>}{{$\gg$}}2
        {||}{{$\|$}}1
        {<=>}{{$\Leftrightarrow$}}2 
        {<->}{{$\leftrightarrow$}}2
        {(+)}{{$\oplus$}}1
        {(-)}{{$\ominus$}}1
}

\lstdefinelanguage{bigIOALang}{%
  basicstyle=\ttfamily,
  keywordstyle=\rmfamily\bfseries,
  identifierstyle=,
  keywords={assumes,automaton,axioms,backward,by,case,choose,components,const,%
    d,det,discrete,do,eff,else,elseif,ensuring,enumeration,evolve,fi,for,forward,from,hidden,if,in%
    input,initially,internal,invariant,local,od,of,output,pre,schedule,signature,so,%
    tasks, simulation,states,stop,tasks,that,then,to,trajdef,trajectories,transitions,tuple,type,union,urgent,uses,when,where,yield},
  literate=
        {\\in}{{$\in$}}1
        {\\preceq}{{$\preceq$}}1
        {\\subset}{{$\subset$}}1
        {\\subseteq}{{$\subseteq$}}1
        {\\supset}{{$\supset$}}1
        {\\supseteq}{{$\supseteq$}}1
        {<}{{$<$}}1
        {>}{{$>$}}1
        {=}{{$=$}}1
        {~}{{$\neg$}}1 
        {|}{{$\mid$}}1
        {'}{{$^\prime$}}1
        {\\A}{{$\forall$}}1 {\\E}{{$\exists$}}1
        {\\/}{{$\vee$}}1 {/\\}{{$\wedge$}}1 
        {=>}{{$\Rightarrow$}}1 
        {->}{{$\rightarrow$}}1 
        {<=}{{$\leq$}}1 {>=}{{$\geq$}}1 {~=}{{$\neq$}}1
        {\\U}{{$\cup$}}1 {\\I}{{$\cap$}}1
        {|-}{{$\vdash$}}1 {-|}{{$\dashv$}}1
        {<<}{{$\ll$}}2 {>>}{{$\gg$}}2
        {||}{{$\|$}}1
        {<=>}{{$\Leftrightarrow$}}2 
        {<->}{{$\leftrightarrow$}}2
        {(+)}{{$\oplus$}}1
        {(-)}{{$\ominus$}}1
}

\lstnewenvironment{BigIOA}%
  {\lstset{language=bigIOALang,basicstyle=\ttfamily}
   \csname lst@SetFirstLabel\endcsname}
  {\csname lst@SaveFirstLabel\endcsname\vspace{-4pt}\noindent}

\lstnewenvironment{SmallIOA}%
  {\lstset{language=ioaLang,basicstyle=\ttfamily\scriptsize}
   \csname lst@SetFirstLabel\endcsname}
  {\csname lst@SaveFirstLabel\endcsname\noindent}

\newcommand{\true}{\mathit{true}}
\newcommand{\false}{\mathit{false}}

\newlength{\bracklen}

\newcommand{\tri}[3]{\ensuremath{\mathit{#1}^\mathit{#2}_\mathit{#3}}}

\newcommand{\sugLocalVars}[2]{\ifthenelse{\equal{}{#2}}%
                             {\tri{localVars}{#1}{desug}}%
                             {\tri{localVars}{#1}{#2,desug}}}
\newcommand{\sugVars}[2]{\ifthenelse{\equal{}{#2}}%
                        {\tri{vars}{#1}{desug}}%
                        {\tri{vars}{#1}{#2,desug}}}

\newenvironment{subSyntax}{\begin{array}{l}}{\end{array}}

\newcommand{\ms}[1]{\ifmmode%
\mathord{\mathcode`-="702D\it #1\mathcode`\-="2200}\else%
$\mathord{\mathcode`-="702D\it #1\mathcode`\-="2200}$\fi}

\def\A{{\cal A}} 
\def\D{{\cal D}} 
\def\T{{\cal T}} 

\newcommand{\vv}{{\bf v}}

\newcommand{\vx}{{\bf x}}

\newcommand{\arrow}[1]{\mathrel{\stackrel{#1}{\rightarrow}}}

\lstdefinelanguage{pvs}{
  basicstyle=\tt \figuresize,
  keywordstyle=\sc \figuresize,
  identifierstyle=\it \figuresize,
  emphstyle=\tt \figuresize,
  mathescape=true,
  tabsize=20,
  sensitive=false,
  columns=fullflexible,
  keepspaces=false,
  flexiblecolumns=true,
  basewidth=0.05em,
  moredelim=[il][\rm]{//},
  moredelim=[is][\sf \figuresize]{!}{!},
  moredelim=[is][\bf \figuresize]{*}{*},
  keywords={and, 
  	 begin,
  	 cases, const,
  	 do,
  	 external, else, exists, end, endcases, endif,
  	 fi,for, forall, from,
  	 hidden,
  	 in, if, importing,
     let, lambda, lemma,
     measure, 
     not,
     or, of,
     return, recursive,
     stop, 
     theory, that,then, type, types, type+, to, theorem,
     var,
     with,while},
  emph={nat, setof, sequence, eq, tuple, map, array, enumeration, bool, real, exp, nnreal, posreal},   
   literate=
        {(}{{$($}}1
        {)}{{$)$}}1
        {\\in}{{$\in\ $}}1
        {\\mapsto}{{$\rightarrow\ $}}1
        {\\preceq}{{$\preceq\ $}}1
        {\\subset}{{$\subset\ $}}1
        {\\subseteq}{{$\subseteq\ $}}1
        {\\supset}{{$\supset\ $}}1
        {\\supseteq}{{$\supseteq\ $}}1
        {\\forall}{{$\forall$}}1
        {\\le}{{$\le\ $}}1
        {\\ge}{{$\ge\ $}}1
        {\\gets}{{$\gets\ $}}1
        {\\cup}{{$\cup\ $}}1
        {\\cap}{{$\cap\ $}}1
        {\\langle}{{$\langle$}}1
        {\\rangle}{{$\rangle$}}1
        {\\exists}{{$\exists\ $}}1
        {\\bot}{{$\bot$}}1
        {\\rip}{{$\rip$}}1
        {\\emptyset}{{$\emptyset$}}1
        {\\notin}{{$\notin\ $}}1
        {\\not\\exists}{{$\not\exists\ $}}1
        {\\ne}{{$\ne\ $}}1
        {\\to}{{$\to\ $}}1
        {\\implies}{{$\implies\ $}}1
        {<}{{$<\ $}}1
        {>}{{$>\ $}}1
        {=}{{$=\ $}}1
        {~}{{$\neg\ $}}1
        {|}{{$\mid$}}1
        {'}{{$^\prime$}}1
        {\\A}{{$\forall\ $}}1
        {\\E}{{$\exists\ $}}1
        {\\/}{{$\vee\,$}}1
        {\\vee}{{$\vee\,$}}1
        {/\\}{{$\wedge\,$}}1
        {\\wedge}{{$\wedge\,$}}1
        {->}{{$\rightarrow\ $}}1
        {=>}{{$\Rightarrow\ $}}1
        {->}{{$\rightarrow\ $}}1
        {<=}{{$\Leftarrow\ $}}1
        {<-}{{$\leftarrow\ $}}1
        {~=}{{$\neq\ $}}1
        {\\U}{{$\cup\ $}}1
        {\\I}{{$\cap\ $}}1
        {|-}{{$\vdash\ $}}1
        {-|}{{$\dashv\ $}}1
        {<<}{{$\ll\ $}}2
        {>>}{{$\gg\ $}}2
        {||}{{$\|$}}1
        {[}{{$[$}}1
        {]}{{$\,]$}}1
        {[[}{{$\langle$}}1
        {]]]}{{$]\rangle$}}1
        {]]}{{$\rangle$}}1
        {<=>}{{$\Leftrightarrow\ $}}2
        {<->}{{$\leftrightarrow\ $}}2
        {(+)}{{$\oplus\ $}}1
        {(-)}{{$\ominus\ $}}1
        {_i}{{$_{i}$}}1
        {_j}{{$_{j}$}}1
        {_{i,j}}{{$_{i,j}$}}3
        {_{j,i}}{{$_{j,i}$}}3
        {_0}{{$_0$}}1
        {_1}{{$_1$}}1
        {_2}{{$_2$}}1
        {_n}{{$_n$}}1
        {_p}{{$_p$}}1
        {_k}{{$_n$}}1
        {-}{{$\ms{-}$}}1
        {@}{{}}0
        {\\delta}{{$\delta$}}1
        {\\R}{{$\R$}}1
        {\\Rplus}{{$\Rplus$}}1
        {\\N}{{$\N$}}1
        {\\times}{{$\times\ $}}1
        {\\tau}{{$\tau$}}1
        {\\alpha}{{$\alpha$}}1
        {\\beta}{{$\beta$}}1
        {\\gamma}{{$\gamma$}}1
        {\\ell}{{$\ell\ $}}1
        {--}{{$-\ $}}1
        {\\TT}{{\hspace{1.5em}}}3        
      }

\lstdefinelanguage{BigPVS}{
  basicstyle=\tt,
  keywordstyle=\sc,
  identifierstyle=\it,
  emphstyle=\tt ,
  mathescape=true,
  tabsize=20,
  sensitive=false,
  columns=fullflexible,
  keepspaces=false,
  flexiblecolumns=true,
  basewidth=0.05em,
  moredelim=[il][\rm]{//},
  moredelim=[is][\sf \figuresize]{!}{!},
  moredelim=[is][\bf \figuresize]{*}{*},
  keywords={and, 
  	 begin,
  	 cases, const,
  	 do, datatype,
  	 external, else, exists, end, endif, endcases,
  	 fi,for, forall, from,
  	 hidden,
  	 in, if, importing,
     let, lambda, lemma,
     measure,
     not,
     or, of,
     return, recursive,
     stop, 
     theory, that,then, type, types, type+, to, theorem,
     var,
     with,while},
  emph={nat, setof, sequence, eq, tuple, map, array, first, rest, add, enumeration, bool, real, posreal, nnreal},   
   literate=
        {(}{{$($}}1
        {)}{{$)$}}1
        {\\in}{{$\in\ $}}1
        {\\mapsto}{{$\rightarrow\ $}}1
        {\\preceq}{{$\preceq\ $}}1
        {\\subset}{{$\subset\ $}}1
        {\\subseteq}{{$\subseteq\ $}}1
        {\\supset}{{$\supset\ $}}1
        {\\supseteq}{{$\supseteq\ $}}1
        {\\forall}{{$\forall$}}1
        {\\le}{{$\le\ $}}1
        {\\ge}{{$\ge\ $}}1
        {\\gets}{{$\gets\ $}}1
        {\\cup}{{$\cup\ $}}1
        {\\cap}{{$\cap\ $}}1
        {\\langle}{{$\langle$}}1
        {\\rangle}{{$\rangle$}}1
        {\\exists}{{$\exists\ $}}1
        {\\bot}{{$\bot$}}1
        {\\rip}{{$\rip$}}1
        {\\emptyset}{{$\emptyset$}}1
        {\\notin}{{$\notin\ $}}1
        {\\not\\exists}{{$\not\exists\ $}}1
        {\\ne}{{$\ne\ $}}1
        {\\to}{{$\to\ $}}1
        {\\implies}{{$\implies\ $}}1
        {<}{{$<\ $}}1
        {>}{{$>\ $}}1
        {=}{{$=\ $}}1
        {~}{{$\neg\ $}}1
        {|}{{$\mid$}}1
        {'}{{$^\prime$}}1
        {\\A}{{$\forall\ $}}1
        {\\E}{{$\exists\ $}}1
        {\\/}{{$\vee\,$}}1
        {\\vee}{{$\vee\,$}}1
        {/\\}{{$\wedge\,$}}1
        {\\wedge}{{$\wedge\,$}}1
        {->}{{$\rightarrow\ $}}1
        {=>}{{$\Rightarrow\ $}}1
        {->}{{$\rightarrow\ $}}1
        {<=}{{$\Leftarrow\ $}}1
        {<-}{{$\leftarrow\ $}}1
        {~=}{{$\neq\ $}}1
        {\\U}{{$\cup\ $}}1
        {\\I}{{$\cap\ $}}1
        {|-}{{$\vdash\ $}}1
        {-|}{{$\dashv\ $}}1
        {<<}{{$\ll\ $}}2
        {>>}{{$\gg\ $}}2
        {||}{{$\|$}}1
        {[}{{$[$}}1
        {]}{{$\,]$}}1
        {[[}{{$\langle$}}1
        {]]]}{{$]\rangle$}}1
        {]]}{{$\rangle$}}1
        {<=>}{{$\Leftrightarrow\ $}}2
        {<->}{{$\leftrightarrow\ $}}2
        {(+)}{{$\oplus\ $}}1
        {(-)}{{$\ominus\ $}}1
        {_i}{{$_{i}$}}1
        {_j}{{$_{j}$}}1
        {_{i,j}}{{$_{i,j}$}}3
        {_{j,i}}{{$_{j,i}$}}3
        {_0}{{$_0$}}1
        {_1}{{$_1$}}1
        {_2}{{$_2$}}1
        {_n}{{$_n$}}1
        {_p}{{$_p$}}1
        {_k}{{$_n$}}1
        {-}{{$\ms{-}$}}1
        {@}{{}}0
        {\\delta}{{$\delta$}}1
        {\\R}{{$\R$}}1
        {\\Rplus}{{$\Rplus$}}1
        {\\N}{{$\N$}}1
        {\\times}{{$\times\ $}}1
        {\\tau}{{$\tau$}}1
        {\\alpha}{{$\alpha$}}1
        {\\beta}{{$\beta$}}1
        {\\gamma}{{$\gamma$}}1
        {\\ell}{{$\ell\ $}}1
        {--}{{$-\ $}}1
        {\\TT}{{\hspace{1.5em}}}3        
      }

\lstdefinelanguage{pvsNums}[]{pvs}
{
  numbers=left,
  numberstyle=\tiny,
  stepnumber=2,
  numbersep=4pt
}

\lstdefinelanguage{pvsNumsRight}[]{pvs}
{
  numbers=right,
  numberstyle=\tiny,
  stepnumber=2,
  numbersep=4pt
}

\lstnewenvironment{BigPVS}%
  {\lstset{language=BigPVS}}
  {}

\lstnewenvironment{PVSNums}%
  {
  \if@firstcolumn
    \lstset{language=pvs, numbers=left, firstnumber=auto}
  \else
    \lstset{language=pvs, numbers=right, firstnumber=auto}
  \fi
  }
  {}

\lstnewenvironment{PVSNumsRight}%
  {
    \lstset{language=pvs, numbers=right, firstnumber=auto}
  }
  {}

\newcommand{\linefigpvs}[9]{

}

\lstdefinelanguage{pvsproof}{
  basicstyle=\tt \figuresize,
  mathescape=true,
  tabsize=4,
  sensitive=false,
  columns=fullflexible,
  keepspaces=false,
  flexiblecolumns=true,
  basewidth=0.05em,
}

\newcommand{\abs}[1]{\left\vert#1\right\vert}

\newcommand{\tuple}[1]{\left\langle#1\right\rangle}

\newcommand{\maxel}[2]{\underset{#2}{\operatorname{max}} #1}

\def\Loc{\act{Loc}}

\newcommand{\toolpassel}{{\it Passel}\xspace}

\newcommand{\dom}{\mathop{\mathsf {dom}}}

\def\dom{{\mathop{\mathsf {dom}}}}

\def\N{\act{N}}
\def\ID{[\N]}

\def\IDE{\ID_{\bot}}

\def\Loc{\act{Loc}}

\newcommand{\val}[1]{{\mathit{val}(#1)}}

\newcommand{\localvar}[2]{{{#1_{#2}}}}

\def\xi{\localvar{x}{i}}

\def\reach{{\sf Reach}}

\newcommand{\nbrleft}[1]{\localvar{L}{#1}}
\newcommand{\nbrright}[1]{\localvar{R}{#1}}

\def\nbrlefti{\nbrleft{i}}
\def\nbrrighti{\nbrright{i}}

\newcommand{\localvarsup}[3]{{{#1_{#2}^{#3}}}}

\def\Vpaneli{\localvarsup{V}{i}{sp}}
\def\Ipaneli{\localvarsup{I}{i}{sp}}
\def\Vdci{\localvarsup{V}{i}{\mathit{dc}}}
\def\Vrefi{\localvarsup{V}{i}{\mathit{ref}}}
\def\Idci{\localvarsup{I}{i}{\mathit{dc}}}
\def\Vaci{\localvarsup{V}{i}{\mathit{ac}}}
\def\Iaci{\localvarsup{I}{i}{\mathit{ac}}}

\def\Vac{V^{\mathit{ac}}}
\def\Vacpeak{V^{\mathit{p}}}
\def\Vgrid{V^{\mathit{grid}}}
\def\Vgridrms{V^{\mathit{rms}}}

\newcommand{\tclock}[1]{\localvarsup{u}{#1}{\mathit{ac}}}
\def\tclocki{\tclock{i}}
\def\tclockdoti{\localvarsup{\dot{u}}{i}{\mathit{ac}}}

\newcommand{\Ahbridge}[1]{\A_{#1}^{ac}}
\def\Ahbridgei{\Ahbridge{i}}
\newcommand{\Abuck}[1]{\A_{#1}^{dc}}
\def\Abucki{\Abuck{i}}

\def\buckClocki{\localvarsup{\tau}{i}{\mathit{dc}}}
\def\dutyBucki{\localvarsup{\delta}{i}{\mathit{dc}}}
\def\switchBucki{\localvarsup{T}{i}{\mathit{dc}}}

\def\twaitiset{\localvarsup{\Delta}{i}{ac}}

\def\twaitzpi{\localvarsup{\delta}{i}{\mathit{z+}}}
\def\twaitpi{\localvarsup{\delta}{i}{\mathit{p}}}
\def\twaitzni{\localvarsup{\delta}{i}{\mathit{z-}}}
\def\twaitni{\localvarsup{\delta}{i}{\mathit{n}}}

\def\Nbrsi{\localvar{\mathit{Nbrs}}{i}}

\def\failedi{\localvar{\mathit{F}}{i}}
\def\failedj{\localvar{\mathit{F}}{j}}
\def\idi{\localvar{\mathit{id}}{i}}

\def\flefti{\localvarsup{\mathit{L}}{i}{F}}
\def\frighti{\localvarsup{\mathit{R}}{i}{F}}

\def\ID{\mathit{ID}}
\def\IDE{\ID_{\bot}}

\newcommand{\IDdef}[1]{\ID_{#1}}
\def\IDF{\IDdef{\act{F}}}
\def\NF{\N_{\act{F}}}
\def\IDNF{\IDdef{\act{O}}}
\def\NNF{\N_{\act{O}}}

\def\Tgrid{T^{ac}}
\def\fgrid{f^{ac}}

\newcommand{\contstate}[1]{#1(t)}

\def\Edg{\mathit{Edg}}
\def\Rst{\mathit{Rst}}

\newcommand{\Grd}{{\mathit{Grd}}}

\newcommand{\Inv}{{\mathit{Inv}}}
\newcommand{\Flow}{{\mathit{Flow}}}

\def\Vi{\mathit{Var_i}}
\def\Xi{\mathit{X_i}}
\def\Qi{Q_i}

\begin{document}
%

\title{\titlename}

%
%
%

\author{\authorluan,~\IEEEmembership{Student Member,~IEEE}
        and~Taylor~T~Johnson,~\IEEEmembership{Member,~IEEE}%
\thanks{L.V. Nguyen and T.T. Johnson are with the Department of Computer Science and Engineering, University of Texas at Arlington, Arlington, TX 76019 USA, e-mail: (luanvnguyen@mavs.uta.edu, taylor.johnson@uta.edu.}
\thanks{Manuscript received March 17, 2014; revised xxx.}}

%
%

\markboth{Transactions on Energy Conversion,~Vol.~X, No.~Y, March~201x}%
{Nguyen \MakeLowercase{\textit{et al.}}: \titlename}
%



\maketitle

\begin{abstract}
%
In this paper, we present the virtual prototyping of a solar array with a grid-tie implemented as a distributed inverter and controlled using distributed algorithms.  Due to the distributed control and inherent redundancy in the array composed of many panels and inverter modules, the virtual prototype exhibits fault-tolerance capabilities.  The distributed identifier algorithm allows the system to keep track of the number of operating panels to appropriately regulate the DC voltage output of the panels using buck-boost converters, and determine appropriate switching times for H-bridges in the grid-tie.  We evaluate the distributed inverter, its control strategy, and fault-tolerance through simulation in Simulink/Stateflow.  Our virtual prototyping framework allows for generating arrays and grid-ties consisting of many panels, and we evaluate arrays of five to dozens of panels.  Our analysis suggests the achievable total harmonic distortion (THD) of the system may allow for operating the array in spite of failures of the power electronics, control software, and other subcomponents.
\end{abstract}


%
\IEEEpeerreviewmaketitle

\ifCLASSOPTIONcaptionsoff
  \newpage
\fi



%
%

\setlength{\belowdisplayskip}{2pt} \setlength{\belowdisplayshortskip}{2pt} 
\setlength{\abovedisplayskip}{2pt} \setlength{\abovedisplayshortskip}{2pt} 

\section{Introduction}
\seclabel{intro}
Multilevel inverters have become popular in recent years for a plethora of reasons, such as their ease of implementation, efficiency, fault-tolerance capabilities, etc.~\cite{lai1996,tolbert1999,mcgrath2002tie,kjaer2005tia,franquelo2008iem,cecati2010tii,johnson_brian2013phdthesis}.
In this paper, we describe the model-based design and virtual prototyping analysis of a grid-tied solar array implemented with fault-tolerant distributed control.
The solar array consists of $\N$ solar panels composed of photovoltaic (PV) modules and corresponding electronics.
Each panel's electronics implement maximum power-point tracking (MPPT) and regulate the panel output voltage using a buck-booster converter.
A $(2\N+1)$-level multilevel inverter is implemented using H-bridges to create a grid-tie.
The control logic for each panel, corresponding buck-booster converter, and H-bridge module is implemented using a separate microcontroller.
An inverter module is the complete plant and computer controller consisting of a panel, its microcontroller, buck-boost converter, etc.
See~\figref{array_architecture} for an overview of the array architecture.

The modules communicate with one another to ensure they switch at appropriate times to create the AC waveform for the grid.
%
%
Next, a distributed identifier service is used by the $\N$ microcontrollers to determine (a) the number of non-faulty modules, and (b) the switching time for each non-faulty module to minimize total harmonic distortion (THD) for the AC grid-tie~\cite{filho2011tia}.
This setup makes the system modular, where it is not necessary to know the number of functioning modules $\NNF \leq \N$, a priori, as the distributed algorithm determines this.
In addition, the distributed identifier service lends the system to be fault-tolerant, whereby if any of the $\N$ panels and corresponding control modules fails, the remaining panels and modules continue operating to ensure the grid-tie remains operational with reasonable THD and response time.
%
%
%
We utilize an abstract failure model, where crash faults of any microcontroller are detected and tolerated, as are actuator stuck-at errors, which corresponds to failed switches in the H-bridges.
We characterize the THD of the system as a function of $\NF$, the number of faulty modules, since as the number of faulty modules increases, the best response of the array will decrease, as the sinusoidal approximation has fewer discrete levels.
In the optimal case, the best achievable THD of an array with $\N$ total modules and $\NF$ faulty modules is that of an array with $\NNF = \N - \NF$ functioning modules.

\begin{figure}[t!]%
	\centering%
    \includegraphics[width=0.8\columnwidth]{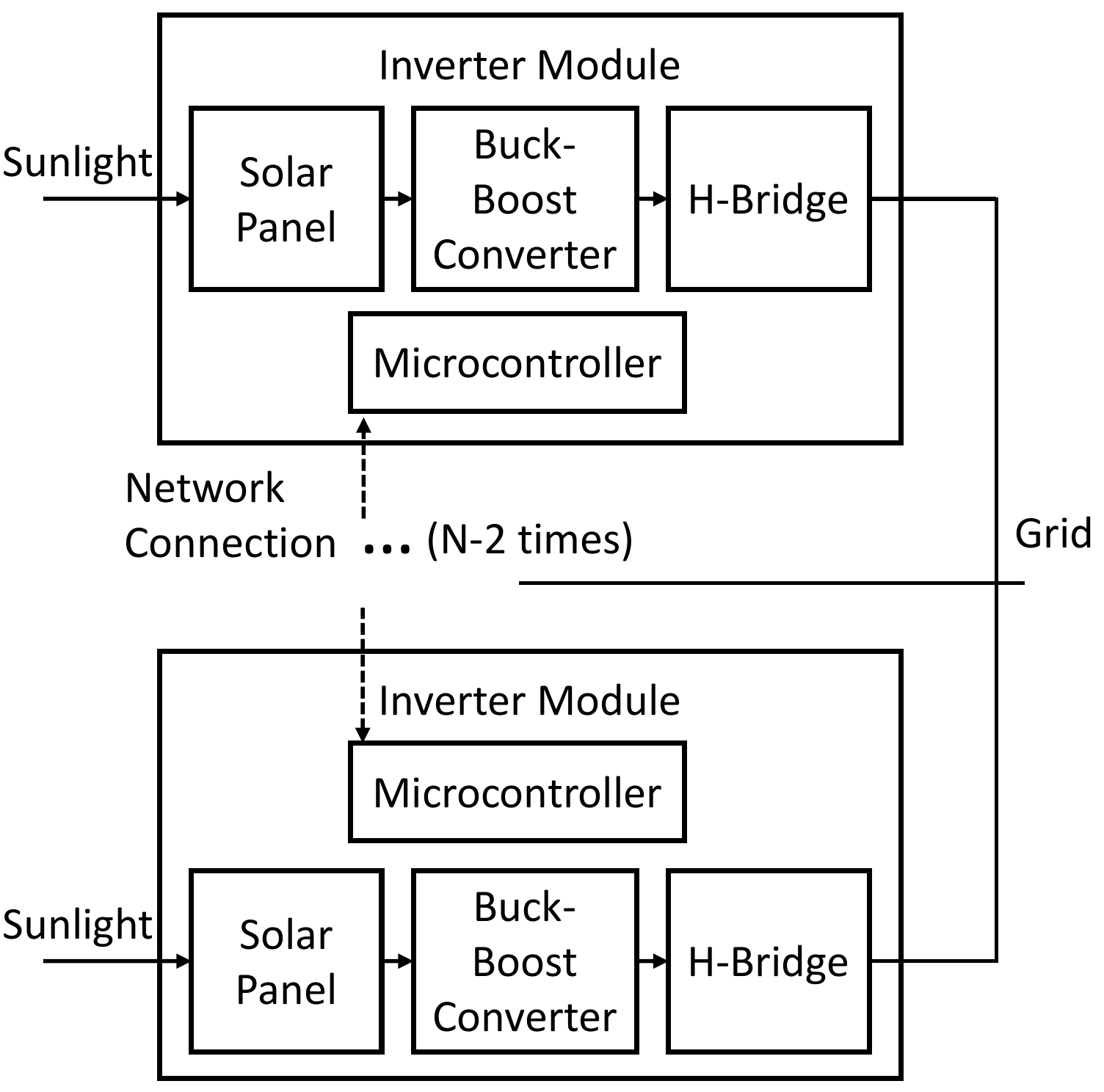}%
	\vspace{-1em}%
	\caption{Overview of the grid-tied solar array, consisting of $\N$ inverter modules, each of which is composed of a solar panel, a microcontroller, a buck-boost converter, and an H-bridge for selecting polarity.}%
	\figlabel{array_architecture}%
	\vspace{-2em}%
\end{figure}%

\paragraph*{Related Work}
There is extensive literature~\cite{turpin2002tie,chen2005tpe,chan2006ciep,ma2007tpe,khomfoi2007tpe,ristow2008tie,lezana2010tie,song2013tpe,harb2013tpe,zhao2013tii,fischer2014tii} regarding fault-tolerance capabilities of single and multi-phase multilevel inverters, as due to their topology, they have inherent redundancy that may be useful for providing fault-tolerance due to switch and other failures.
For a recent overview of general reliability and fault-tolerance in power electronics, see~\cite{song2013tpe}, for a particular focus on multilevel inverters, see~\cite{lezana2010tie}, and for a focus on the reliability of DC-to-DC converters in PV energy conversion systems, see~\cite{dhople2012tpe}.
In~\cite{turpin2002tie} the reliability of multilevel inverters was studied to present an argument against reliability necessarily decreasing due to increased component counts, each with their own failure rates.
A single-phase fault-tolerant multilevel inverter is developed and experimentally validated with $5$-level prototype in~\cite{chen2005tpe} and focuses on utilizing redundant circuitry and appropriate control for maintaining the output voltage.
For example, Fault-tolerance in multilevel inverters can be achieved by adding some power device to the basic topologies such as fourth-leg~\cite{bolognani2000experimental} or reconfiguring the flying capacitor multilevel inverter into a full binary combination scheme, and balance capacitor voltage  by using three-phase joint switching states~\cite{kou2002full}.
A comparison of several inverter topologies along with their cost and reliability tradeoffs is presented in~\cite{chan2006ciep}.
%
%
In~\cite{ma2007tpe}, a strategy is developed for reconfiguring carrier-based modulation signals to provide fault-tolerance in multilevel inverters due to switches either failing open circuit or short circuit, and is experimentally evaluated on a three-phase five-level prototype.
The authors of~\cite{khomfoi2007tpe} develop a fault diagnosis system for multilevel inverters using neural networks.
%
%
%

Overall, the vast majority of fault-tolerance capabilities in multilevel inverters focus on handling hardware faults using redundant hardware and topology (\ie physical) solutions.
In contrast, in this paper, we consider primarily software-based fault-tolerance methods and have the capability to handle both hardware (\eg switch failures) and software faults (\eg microcontroller crashes) using software (\ie cyber) solutions.
The topology of the inverter we consider in this paper is very similar to that of~\cite{johnson2011apec,johnson_brian2013phdthesis}, but we utilize a buck-boost converter for DC voltage regulation, and we focus on distributed control instead of communication-less control.
We do not focus on any particular maximum power point tracking (MPPT) scheme in this paper, but refer readers to numerous methods and their tradeoffs in~\cite{esram2007tec}.

%


%
%

Our array simulator is developed in Simulink/Stateflow, and similar simulation models have been developed previously~\cite{patel2008tec,ropp2009tec,ding2012tec,maki2013tec}.
A MATLAB simulation model for PV modules is presented in~\cite{patel2008tec} and considers factors such as temperature, shading, etc.
In~\cite{ropp2009tec}, the authors develop a MATLAB/Simulink model of a grid-connected single-phase array with MPPT, but do not consider multilevel inverters as we do.
In~\cite{ding2012tec}, the authors develop a MATLAB/Simulink model of PV modules accounting for numerous non-idealities, such as nonuniform irradiance.
A detailed MATLAB/Simulink for studying partial shading of arrays is studied in~\cite{maki2013tec}.



\paragraph*{Contributions}
The main contributions of this paper are: (a) the development and implementation of the fault-tolerant distributed control strategy for solar-to-AC conversion, (b) the holistic design and analysis of a cyber-physical system (CPS) in a virtual prototyping environment (MATLAB/Simulink/Stateflow), and (c) the application of hybrid systems modeling techniques for virtual prototyping.
We highlight that in contrast to most existing work on fault-tolerance of multilevel inverters, the failure model considered in this paper is an abstraction of both cyber and physical failures, and works by coordination through distributed control.

\paragraph*{Paper Organization}
The remainder of this paper is organized as follows.
\secref{model} presents the distributed solar array architecture and its control, including the communication and computation capabilities of its subcomponents, as well as a failure model of the subcomponents.
%
%
\secref{simulation} presents the simulation-based analysis of the virtual prototype, including comparisons of THD with and without failures, different failure modes, and arrays consisting of $\N = 5$ to $\N = 35$ panels.
\secref{conclusion} concludes the paper and presents directions for future work.


\section{Distributed Array Architecture and Modeling}
\seclabel{model}

\paragraph*{Preliminaries}
For a set $S$, let $\abs{S}$ be the cardinality of $S$, which is the number of elements in $S$.
For a set $S$, let $S_{\bot}$ be $S \cup \{ \bot \}$ where $\bot \notin S$.
We model several of the cyber-physical components of the array using the hybrid automaton formalisms, and refer interested readers to~\cite{alur1995tcs,lynch2003ic,kaynar2006book,johnson2013phdthesis} for detailed definitions of such modeling formalisms, and to~\cite{johnson2012peci,hossain2013peci,nguyen2014arch,althoff2014tps} for descriptions specified to power electronics and systems.
We begin by briefly reviewing hybrid automata.
A \emph{hybrid automaton} is a (possibly nondeterministic) state machine with state that can evolve both instantaneously (through {\em discrete transitions\/}) and over intervals of time (according to {\em trajectories\/}).
Variables are associated with types and are used as names for state components, such as currents, voltages, and times.
For a set of variables $V$, a valuation $\vv$ is a function that maps each variable $v \in V$ to a point in its type, denoted $\type{v}$.
The set of all possible valuations is $\val{V}$.
For a valuation $\vx$, we use $\vx.x$ to denote the value of the variable $x \in V$.
Since the distributed system is composed of $\N$ panels, each of which has its own power electronics, software, etc., we model the $i^{th}$ panel as an automaton $\A_i$.

Mathematically, a hybrid automaton $\A_i$ is a tuple $\tuple{\Vi, \Loc_i, \Qi, \Theta_i, \Edg_i, \Grd_i, \Rst_i, \Flow_i, \Inv_i}$, where:
\begin{inparaenum}[(a)]
	\item $\Vi$: is a set of variables, where $\Xi \subseteq \Vi$ are the continuous, real-typed variables.
	\item $\Loc_i$: is a set of discrete locations.
	\item $\Qi \deq \val{\Vi}$ is the set of states, and is the set of all valuations of each variable $v \in \Vi$.
	A {\em state\/} is denoted by bold $\vx$ and assigns values to every variable in the set of variables $\Vi$.
	For a state $\vx \in \Qi$, the valuation of $\vx.loc$ is called the {\em location\/}, and along with the valuations of any discrete variables, it describes the discrete state.
	The valuation of the continuous variables in $\Xi$, that is $\{\vx.x : x \in \Xi \}$, is called the {\em continuous state\/} and is referred to as $\vx.\Xi$.
\item $\Theta_i \subseteq \Qi$ is a set of {\em initial states\/}.
\item $\Edg_i$ is the set of {\em edges\/}.
\item $\Grd_i : \Edg_i \arrow{} \Qi$ is a function that associates a {\em guard\/} (a valuation of $V$ that must be satisfied such that a transition may be taken) with each edge.
\item $\Rst_i : \Edg_i \arrow{} (\Qi \arrow{} 2^{\Qi})$ is a function, called the {\em reset map\/}, associated with each edge.
A reset map associates a set of states with each edge.
\item $\Flow_i : \Loc_i \arrow{} (\Qi \arrow{} 2^{\Qi})$ associates a {\em flow map\/} with each location.
\item $\Inv_i: \Loc_i \arrow{} 2^{\Qi}$ associates an {\em invariant\/} with each location.
\end{inparaenum}


The semantics of $\A_i$ are defined in terms of sets of {\em transitions\/} and {\em trajectories\/}.
The set of transitions $\D_i \subseteq \Qi \times \Qi$ is defined as follows.
We have $(\vv, \vv') \in \D_i$ if and only if, for $e = (\vv.loc, \vv'.loc)$,
\begin{inparaenum}[(a)]
\item $e \in \Edg_i$,
\item $\vv \in \Grd_i(e)$, and 
\item $\vv' \in \Rst_i(e)(\vv.X)$.
\end{inparaenum}
A {\em trajectory\/} for $\A_i$ is a function $\tau:[0,t] \arrow{} Q_i$ that maps an interval of time to states such that:
\begin{inparaenum}[(a)]
\item For all $t' \in [0,t]$, $\tau(t').loc = \tau(0).loc$, that is, the discrete state remains constant,
\item $(\tau \restrrange X)$, that is, the restriction of $\tau$ to $X_i$ is a solution of the differential equation specified by the flow function $\dot{X}_i = \Flow_i(\tau(0).loc)(\tau(0))$, and
\item For all $t' \in [0, t]$, $\tau(t') \in \Inv_i(\tau(0).loc)$.
\end{inparaenum}
The set of all the trajectories of $\A_i$ is written $\T_i$.
The domain for a trajectory $\tau \in \T_i$ is denoted by $\tau.\dom$.
We define $\tau.\ltime$ as the right endpoint of $\tau.\dom$, $\tau.\lstate \deq \tau(\tau.\ltime)$, and $\tau.\fstate \deq \tau(0)$.
An {\em execution\/} of $\A_i$ is a sequence $\alpha = \tau_0 \tau_1 \ldots$, such that:
\begin{inparaenum}[(a)]
\item each $\tau_k \in \T_i$, 
\item for each $k$, $(\tau_k(t), \tau_{k+1}(0)) \in \D_i$, where $t$ is the right endpoint of the domain of $\tau_k$, and 
\item $\tau_0 \in \Theta_i$.
\end{inparaenum}
%
%
A state $\vv \in \Qi$ is said to be {\em reachable\/} if there exists a finite execution $\alpha$ that ends with $\vv$.
%
%
%

%
%
%
%
%

\subsection{Architecture and Modeling}
The distributed solar array consists of $\N$ solar panels and corresponding electronics for implementing the grid-tie (see~\figref{array_architecture}).
For each solar panel, there is also an \emph{inverter module} consisting of a computer, communications system, and power electronics.
Each inverter module's power electronics consist of a DC-to-DC buck-boost converter for regulating the panel's output voltage, and an H-bridge for connecting and disconnecting the panel's output voltage at appropriate times to generate the AC waveform (see~\figref{circuit_array}).
We refer to each panel and its corresponding inverter module as an \emph{agent} with a unique identifier $i \in \ID$, where $\ID \deq \{1, \ldots, \N\}$.
We model the $i^{th}$ solar panel's buck-boost converter as a hybrid automaton (see~\figref{automaton_buckboost}) denoted $\Abucki$, and its H-bridge as a hybrid automaton (see~\figref{automaton_hbridge}) denoted $\Ahbridgei$.
Each panel and inverter module is specified as a hybrid automaton consisting of the composition of the individual components:
\begin{align}
\A_i \ \deq \ \Abuck{i} \ \| \ \Ahbridge{i}.
\end{align}
For a given $\N$, the complete system $\A$ composed of the $\N$ solar panels, $\N$ buck-boost converters, $\N$ H-bridges, and computer control software and hardware is:
\begin{align}
\A \ \deq \ \A_1 \ \| \ \ldots \ \| \ \A_{\N}, \eqlabel{system_composed}
\end{align}
where $\|$ is a parallel (concurrent) composition of automata (see, e.g.,~\cite[Chapter 2]{johnson2013phdthesis}).

\begin{figure}[t!]
	\centering
	\includegraphics[width=0.9\columnwidth]{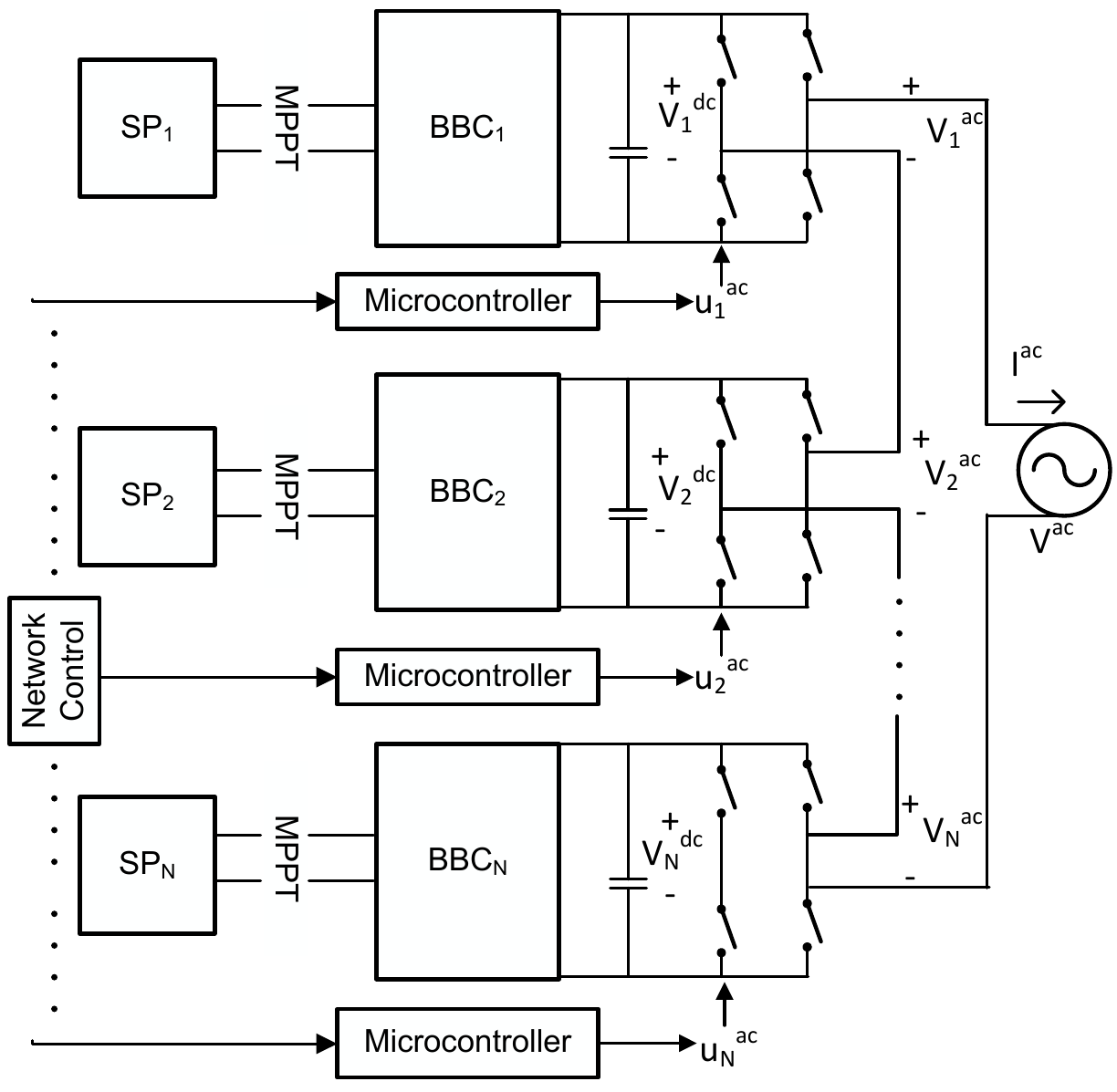}
	\caption{High-level circuit diagram of the array and grid-tie illustrating the solar panels ($SP_i$) controlled with MPPT that feed the panel output voltage $\Vpaneli$ into the buck-boost converters ($BBC_i$) with output voltage $\Vdci$.  Next, the H-bridges switch at appropriate times to connect the $\N$ DC regulated voltage sources $\Vdci$ with potentially reversed polarity in series to create the grid connection voltage $\Vac$.  The buck-boost converter control ($\Vrefi$) and H-bridge switching control ($\tclocki$) for inverter module $i$ depends upon network information from other inverter modules in the array.}
	\figlabel{circuit_array}
	\vspace{-1em}
\end{figure}

Each agent $i \in \ID$ is associated with the following electrical (physical) real variables:
\begin{inparaenum}[(a)]
\item $\Vpaneli$: the voltage output of agent $i$'s solar panel and input to agent $i$'s DC-to-DC converter,
\item $\Ipaneli$: the output current of agent $i$'s solar panel and input to agent $i$'s DC-to-DC converter,
\item $\Vrefi$: the reference voltage for agent $i$'s DC-to-DC converter to track,
\item $\Vdci$: the voltage output of agent $i$'s DC-to-DC converter and input to agent $i$'s H-bridge,
\item $\Idci$: the current output of agent $i$'s DC-to-DC converter and input to agent $i$'s H-bridge,
\item $\Vaci$: the voltage output of agent $i$'s H-bridge and input to the grid, and
\item $\Iaci$: the current output of agent $i$'s H-bridge and input to the grid.
\end{inparaenum}
Additionally, each agent $i \in \ID$ is associated with the following communications and computational (cyber) quantities:
\begin{inparaenum}[(a)]
\item $\twaitiset$ $\deq \{\twaitzpi$, $\twaitpi$, $\twaitzni$, $\twaitzni \}$: a set of switching times for agent $i$'s H-bridge to connect/disconnect $\Vaci$ with what polarity to the grid, 
\item $\tclocki$: the H-bridge control timer for agent $i$ used to compare to the switching times in $\twaitiset$,
\item $\Nbrsi$: the communication neighbors of agent $i$, consisting of the agents to its left (denoted $\nbrlefti$) and right (denoted $\nbrrighti$).
The left and right neighbors are defined to be the adjacent panels, e.g., in~\figref{array_architecture}.
Without failures, we have $\nbrlefti = i - 1$ and $\nbrrighti = i + 1$, for $i \geq 2$ and $i \leq \N - 1$, respectively, but will redefine these in the case of failures shortly.
%
%
\end{inparaenum}
These variables define the set of variables $\Vi$ of the automata $\Abucki$ and $\Ahbridgei$.
As we consider their compositions, we do not differentiate between variables of the two automata.
Additionally, we note that all these variables are mappings from time to elements in the variables' types.
For some $v \in \Vi$, we will denote this interchangeably by $\vx.v$ for some reachable state $\vx$, or by $\contstate{v}$ for some time $t \in \mathbb{R}_{\geq 0}$ such that $t = \tau.\ltime$ and $\vx = \tau.\lstate$, \ie $t$ is the endpoint of a trajectory $\tau$ ending in reachable state $\vx$.


\subsection{Failure Model and Distributed Notification}
We utilize the following failure model of each agent's physical and cyber components, inspired by similar models developed in~\cite{johnson2010icdcs,johnson2011jnsa}.
While H-bridge failure modes could potentially turn them into open circuits, thus disconnecting the array from the grid, we do not consider such scenarios and assume if the H-bridge fails, it fails as a short adding zero voltage to $\Vac$.
We model general abstracted failures of the entire inverter module that do not cause open circuits, such as the microcontroller crashing, the buck-boost converter entering a failure mode, etc.
We assume we have a method to detect failures, e.g., through a heartbeat service for crash failures.
This assumption is reasonable as our primary focus is on cyber failures---e.g., computer crashes and may recover, communication link is lost temporarily, but desire the grid-tie to recover when the computer restarts or the communication link is restored.
Thus, this failure model is an abstraction of more detailed failures.
%
%
Each agent $i \in \ID$ is augmented with an additional Boolean-valued variable $\failedi$ indicating whether it has failed ($\true$) or not ($\false$).
If agent $i \in \ID$ is failed, then $\contstate{\failedi} = \true$, and if not, $\contstate{\failedi} = \false$.
%
The set of failed agents is denoted by $\contstate{\IDF} \subseteq \ID$ and is the set $\{ i \in \ID \ | \ \contstate{\failedi} \}$.
We define the number of failed agents as $\contstate{\NF} \deq \abs{\contstate{\IDF}}$.
%
The set of operating (non-failed) agents is denoted by $\contstate{\IDNF} \subseteq \ID$ and is the set $\ID \setminus \contstate{\IDF}$.
We also define the number of operating agents as $\contstate{\NNF} \deq \abs{\contstate{\IDNF}}$ and we note $\contstate{\NNF} = \N - \contstate{\NF}$.

We assume failures may be detected---e.g., through use of a heartbeat service for computer/software crash failures---and focus on tolerating failures through software as they become known.
A distributed gossip protocol~\cite{lynch1996book} spreads the identifiers of any failed agents throughout the array, so any agent knows within a short period of time if any other agent is failed or not.
Using this information, the left and right neighbors are redefined, respectively, as $\contstate{\nbrlefti} = \max{ \{ j \in \ID | \contstate{\failedj} \wedge j < i \} }$  and $\contstate{\nbrrighti} = \min{ \{ j \in \ID | \contstate{\failedj} \wedge j > i \} }$.

\paragraph*{Distributed Identification and Notification}
Each agent $i \in \ID$ is augmented with a variable $\idi$ with index type ($\type{\idi} = \IDE$), which indicates its identifier in the set of operational agents, $\IDNF$.
First, each agent keeps track of the number of failures to its left (lower identifiers) as $\contstate{\flefti} = \abs{\{ j \in \ID \ | \ \contstate{\failedj} = \true \wedge j < i \}  }$, and symmetrically $\contstate{\frighti}$ for agents to its right (higher identifiers).
We observe that $\contstate{\NF} = \contstate{\flefti} + \contstate{\frighti}$, so agents may compute the number of failed agents.
Each operational agent $i \in \IDNF$ determines $\idi$ using the following local method:
\begin{align}
%
\contstate{\idi} = i - \contstate{\flefti}.
%
\end{align}
Using this method, we have that $\maxel{\contstate{\idi}}{i \in \ID} = \contstate{\NNF}$.
Together, these distributed identifier services allow each operational agent $i \in \IDNF$ to compute the number of operational and failed agents for use in determining the DC voltage reference $\Vrefi$ and switching times $\twaitiset$ as described next.





%

\subsection{Buck-Boost Converter Model and Control}
For the buck-boost converter model, we utilize a hybrid automaton model developed and analyzed in~\cite{nguyen2014arch}.
Each inverter module's buck-boost converter has two real-valued state variables modeling physical quantities: the inductor current $\Idci$ and the capacitor voltage $\Vdci$, depicted in~\figref{circuit_buckboost}.
These two state variables at time $t$ are written in vector form as:
\begin{align*}
\contstate{\xi} = \left[ \begin{array}{c} \contstate{\Idci} \\ \contstate{\Vdci} \end{array} \right].
\end{align*}
We consider a state-space model without the discontinuous conduction mode (DCM), see \eg~\cite{severns1985book,erickson2004book}.

The reference voltage for each DC-to-DC converter is:
\begin{align}
\contstate{\Vrefi} \ \deq \ \frac{\Vacpeak}{\contstate{\NNF}}, \eqlabel{vrefi}
\end{align}
where $\Vacpeak$ is the AC peak voltage (e.g., $\Vacpeak = \sqrt{2} \Vgridrms$ for the root mean square (RMS) AC voltage $\Vgridrms$).
If $\contstate{\Vrefi} < \contstate{\Vpaneli}$, then the buck-boost converter is in a buck mode and decreases its output voltage $\contstate{\Vdci}$.
Otherwise, if $\contstate{\Vrefi} > \contstate{\Vpaneli}$, then the buck-boost converter is in a boost mode and increases its output voltage $\contstate{\Vdci}$.
Note that since $\contstate{\Vrefi}$ is defined in terms of the number of operating agents $\contstate{\NNF}$, it may vary over time.

\begin{figure}[t!]%
	\centering%
\begin{circuitikz} \draw
 (0,0) 	to[V,v=$\Vpaneli$ ] (0,2) 
				to[short,-o](0.75,2);
	\draw[ very thick](0.78,2)-- +(30:0.46);
	\draw (1.25,2)to[short,o-](2,2)
				to[L = $L_i$,*-*,i_=$\Idci$] (2,0) -- (1.75,0)
				to (0,0)
	(4,2) to[D*,*-*] (2,2)
	(4,2) to[C = $C_i$,*-*,v_=$V_i^C$] (4,0)
	(4,2) to (6,2) 
	(6,2) to[R = $R_i$,v_=$\Vdci$] (6,0)
				to (0,0) ;
\end{circuitikz}%
	\caption{Buck-boost converter circuit---a DC input $\Vpaneli$ is increased or decreased to a higher or lower DC output $\Vdci$.}%
	\figlabel{circuit_buckboost}%
	\vspace{-1em}
\end{figure}
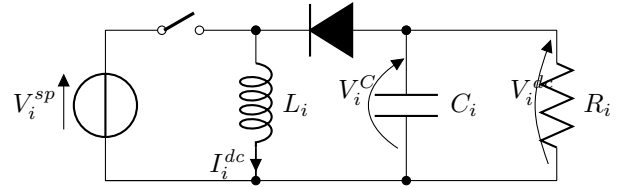

\begin{table}[tb]%
	\centering%
	\footnotesize%
	\bgroup%
	\begin{tabular}{|l|c|c|c|}
		\hline
		Switch $S_i$ State & $A_i^m$ & $B_i^m$ & Duty Cycle $\contstate{\dutyBucki}$ \\
		\hline
		\rule{0pt}{5ex} Open &  $\left[ \begin{array}{cc} 0 & -\frac{1}{L_i} \\ \frac{1}{C_i} & -\frac{1}{R_i C_i} \end{array} \right]$ & $\left[ \begin{matrix} 0\\ 0 \end{matrix} \right]$   & $\frac{\contstate{\Vrefi}}{\contstate{\Vrefi} + \contstate{\Vpaneli}}$ \\
		\hdashline
		\rule{0pt}{5ex}			Close &  $\left[ \begin{array}{cc} 0 & 0 \\ 0 &-\frac{1}{R_i C_i} \end{array} \right]$   & $\left[ \begin{matrix} \frac{1}{L_i}\\ 0 \end{matrix} \right]$ & \\
		\hline
		\hline
	\end{tabular}%
	\egroup%
	\vspace{0.1cm}%
	\caption{Dynamics of agent $i$'s buck-boost converter $\Abucki$.}%
	\tablabel{dynamics}%
	\vspace{-3em}%
\end{table}%
\normalsize

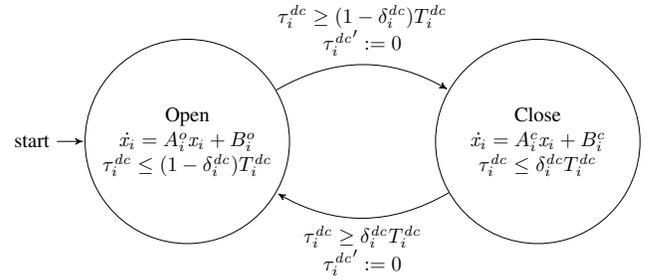
\begin{figure}[t!]%
	\centering%
	\begin{adjustbox}{max size={0.95\columnwidth}{0.75\textheight}}%
	\begin{tikzpicture}[>=stealth',shorten >=1pt,auto,node distance=6cm,font=\normalsize]
		\tikzstyle{every state}=[minimum size=3.5cm,font=\normalsize]
		\node	[initial,state]		(open) 	{\makecell[c]{$\text{Open}$\\$\dot{\mathit{x_i}} = A_i^{o}\mathit{x_i} +  B_i^{o}$\\$\buckClocki \leq (1-\dutyBucki) \switchBucki$}};
		\node[state] (closed)      	[right of=open]		{\makecell[c]{$\text{Close}$\\$\dot{\mathit{x_i}} = A_i^{c}\mathit{x_i} +  B_i^{c}$\\$\buckClocki \leq \dutyBucki \switchBucki $}};
		\path[->]			(closed)	edge[bend left] node{\makecell[c]{$\buckClocki \geq \dutyBucki \switchBucki$\\$\buckClocki' := 0$}} (open); 
		\path[->]			(open)	edge[bend left] node{\makecell[c]{$\buckClocki \geq (1-\dutyBucki) \switchBucki$\\$\buckClocki' := 0$}} (closed); 
	\end{tikzpicture}%
	\end{adjustbox}%
	\caption{Hybrid automaton model $\Abucki$ for agent $i$'s buck-boost converter.  The matrices and vectors $A_i^o$, $A_i^c$, $B_i^o$, and $B_i^c$ are constant but may vary between panels, and $\switchBucki$ is constant.  The state vector $\xi$, duty cycle $\dutyBucki$, and $\buckClocki$ are variables and vary with time.}%
	\figlabel{automaton_buckboost}%
	\vspace{-2em}%
\end{figure}%

\subsection{H-Bridge Modeling and Control}
We model the H-bridge plant as ideal switches, with the controller that connects the output voltage as shown in~\figref{automaton_hbridge} as either:
\begin{inparaenum}[(a)]
\item $\Vaci = 0$: disconnected (locations $Zero^+$ and $Zero^-$),
\item $\Vaci = \Vdci$: connected in series with positive polarity (location $Positive$), or
\item $\Vaci = -\Vdci$: connected in series with reverse polarity (location $Negative$).
\end{inparaenum}
The grid-tie AC voltage $\Vac$ is then defined as the series connection of all $\NNF$ operating inverter modules output voltages:
\begin{align}
\contstate{\Vac} \ \deq \ \sum_{i \in \contstate{\IDNF}} \ \contstate{\Vaci}. \eqlabel{vac}
\end{align}
\begin{figure}[t!]
	\centering
	\includegraphics[width=0.85\columnwidth]{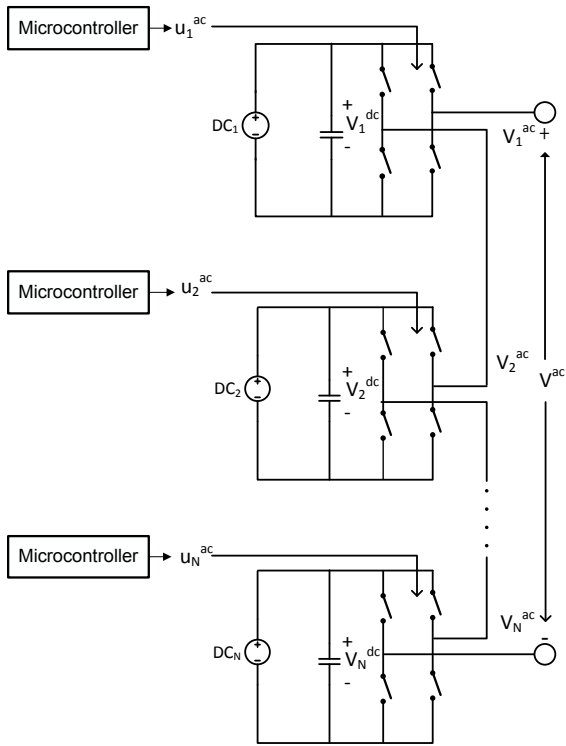}
	\caption{For the purpose of the H-bridge control and finding the switching signals $\tclock{1}$, $\ldots$, $\tclock{\N}$, the panel and buck/boost converter are abstracted and treated as ideal voltage sources ($DC_1$, $\ldots$, $DC_{\N}$).}
	\figlabel{circuit_hbridge}
\end{figure}

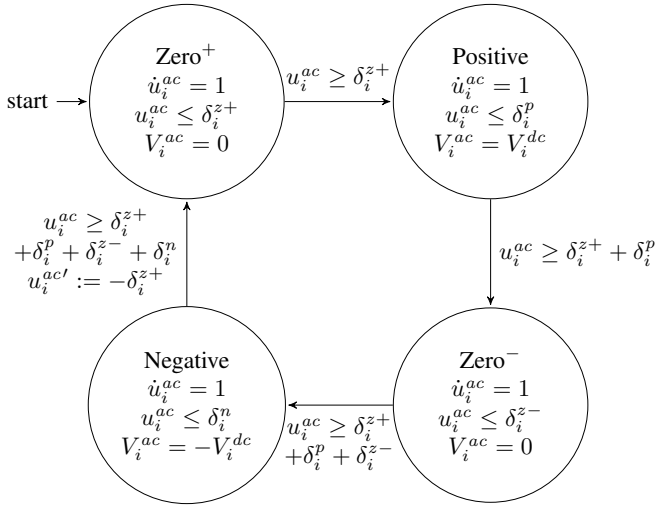
\begin{figure}[t!]
	\vspace{-1em}
	\centering
	\begin{adjustbox}{max size={\columnwidth}{0.75\textheight}}
	\begin{tikzpicture}[>=stealth',shorten >=1pt,auto,node distance=4.45cm]
		\tikzset{every state/.style={minimum size=2.85cm}}
		\node[initial,state] (zp)      			{\makecell[c]{$\text{Zero}^{+}$\\$\tclockdoti = 1$\\$\tclocki \leq \twaitzpi$\\$\Vaci = 0$}};
		\node[state]		 (p) [right of=zp]	{\makecell[c]{$\text{Positive}$\\$\tclockdoti = 1$\\$\tclocki \leq \twaitpi$\\$\Vaci = \Vdci$}}; 
		\node[state]		 (zn) [below of=p]	{\makecell[c]{$\text{Zero}^{-}$\\$\tclockdoti = 1$\\$\tclocki \leq \twaitzni$\\$\Vaci = 0$}};
		\node[state]		 (n) [below of = zp]{\makecell[c]{$\text{Negative}$\\$\tclockdoti = 1$\\$\tclocki \leq \twaitni$\\$\Vaci = -\Vdci$}}; 
		\path[->]			(zp)	edge node{\makecell[c]{$\tclocki \geq \twaitzpi $}} (p);  
		\path[->]			(p)	edge 		node {\makecell[c]{$\tclocki \geq \twaitzpi + \twaitpi $}} (zn); 
		\path[->]			(zn)	edge 	node {\makecell[c]{$\tclocki \geq \twaitzpi$\\$ +\twaitpi + \twaitzni $}} (n); 
		\path[->]			(n)	edge 		node {\makecell[c]{$\tclocki \geq \twaitzpi$\\$ + \twaitpi + \twaitzni + \twaitni $\\$\tclocki' := -\twaitzpi$}} (zp); 
	\end{tikzpicture}
	\end{adjustbox}
	\caption{Hybrid automaton model $\Ahbridgei$ for agent $i$'s H-bridge switching logic.}
	\figlabel{automaton_hbridge}
	\vspace{-2em}
\end{figure}

The set of switching times for the H-bridge to connect $\Vdci$ with different polarities to create $\Vaci$ is denoted:
\begin{align}
\contstate{\twaitiset} \ \deq \ \{ \contstate{\twaitzpi}, \contstate{\twaitpi}, \contstate{\twaitzni}, \contstate{\twaitni} \}, \eqlabel{twaitiset}
\end{align}
where the elements are respectively the time to spend with $\Vaci = 0$, then the time to spend with $\Vaci = \Vdci$, then the time to spend with $\Vaci = 0$ again, and finally the time to spend with $\Vaci = -\Vdci$ before repeating.
See~\figref{execution_vdc_failure} for an example of the switching signals illustrating these various transitions.
For finding the switching times of the H-bridge, we utilize the following protocol and we derive the idealized switching times for each agent $i \in \ID$:
%
%
%
\begin{align*}
\frac{i}{\NNF + 1} \ = \ & \sin\left( \frac{2 \pi t}{\Tgrid} \right), \ \text{and solving for } t, \\
t \ = \ & \frac{\Tgrid}{2 \pi} \sin^{-1}\left( \frac{i}{N+1} \right).
\end{align*}
Of course, $t$ is not unique, but defines the amount of time $\twaitzpi$ spent in the zero state before switching to the positive output state.
The other waiting times simply subdivide the period, and accounting for failures using $i$'s identifier $\idi$ out of the $\NNF$ operating agents, we have:
\begin{align}
\contstate{\twaitzpi} \ = \ & \frac{\Tgrid}{2 \pi} \sin^{-1}\left( \frac{\contstate{\idi}}{\contstate{\NNF} + 1} \right), \eqlabel{hbridge_switch_local}
\end{align}
and likewise for the shifted switching times $\twaitpi$, $\twaitni$, and $\twaitzni$.
We assume that the sinusoid used to generate the switching times in~\eqref{hbridge_switch_local} is synchronized with the grid phase, using, e.g., a phase-locked loop (PLL), which can be implemented in a distributed fashion by informing all operational agents of the grid phase.
Refer to~\figref{execution_vdc_failure} for examples of the switching times generated using this method with failures.

\section{Virtual Prototype Simulation Analysis}
\seclabel{simulation}
Next we describe the simulation setup and analysis of the distributed solar array and inverter virtual prototype.
We wrote a MATLAB program to programmatically generate Simulink/Stateflow (SLSF) models of the array for varying the number of panels and inverter modules ($\N$).
Specifically, for a given $\N$, the program generates an array $\A$ consisting of a panel, inverter module, and its control software composed together, e.g.,~\eqref{system_composed}.
That is, the simulator generates SLSF simulation models corresponding to~\figreftwo{array_architecture}{circuit_array}.
The various parameters used for the circuit components are summarized in~\tabref{vars}.
The grid-tie was configured for a standard residential-style connection at $120$ V and $60$ Hz.
The control logic for both automata $\Abucki$ and $\Ahbridgei$ are implemented as continuous-time state-machines using Stateflow.
Using these programmatically-generated array models, we have performed thousands of simulations for analyzing the system in scenarios with and without failures, as detailed next.
%

%
\begin{table}[tb]
	\footnotesize
	\centering
	\begin{tabular}{|l|c|l|}
		\hline
		Component / Parameter Name & Symbol & Value  																		\\
		\hline
		Buck-Boost Input Voltage					& $\contstate{\Vpaneli}$				&	$18.6$ V $\pm \ \epsilon$				\\
		Desired Buck-Boost Output Voltage	& $\contstate{\Vrefi}$					&	$\frac{\Vgridrms}{\contstate{\NNF}}$ V				\\
		Actual Buck-Boost Output Voltage	& $\contstate{\Vdci}$						&	varies			\\
		Load Resistance										& $R_i$								&	$4$ $\Omega$ $\ \pm \ 5\%$			\\
		Capacitor													& $C_i$								&	$60$ uF $\ \pm \ 5\%$				\\
		Inductor													& $L_i$								&	$40$ uH $\ \pm \ 5\%$					\\
		Switching Period									& $\switchBucki$		&	$4$ $\mu$s					\\
		Switch-closed duty cycle					& $\contstate{\dutyBucki}$			& varies 										\\
		Switch-open duty cycle					  & $1-\contstate{\dutyBucki}$		& varies											\\
		Grid Period												& $\Tgrid$					& $0.0167$ s 	\\
		Grid Frequency										& $\fgrid$					& $60$ Hz 	\\
		Desired Grid Voltage							& $\Vgrid$						& $120$ V$^{rms}$, $60$ Hz	\\
		Actual Array Voltage							  & $\contstate{\Vac}$					& varies	\\
		\hline
	\end{tabular}
	\vspace{0.1cm}
	\caption{Summary of variables and parameters used in simulations.}
	\tablabel{vars}
	\vspace{-2em}
\end{table}

\subsection{Total Harmonic Distortion (THD) with Static Failures}
Static failures are those that occur before the grid-tie is connected and do not affect the dynamic performance.
\figref{execution_vac} shows an example execution for $\N = 35$ panels with both no failures and $\NF = 5$ static failures, along with an execution for $\N = 10$ panels with no failures.
\figref{thd_failure} shows the THD of the array as a function of the number of operating agents, $\NNF$.
Additionally,~\figref{thd_failure} shows the THD for \emph{static} failures, which are those where some agents are failed at start-up and remain failed.
The results illustrate that increasing the number of static failures returns the array to the achievable THD in an array with $\NF$ fewer panels.
The different curves in~\figref{thd_failure} correspond to the numbers of non-failed agents $\NNF$ for a given array of $\N$ panels and inverter modules.
The simulations varied $\NF$ from $0$ (no failures) to $6$ (six failed agents), and $\N$ from $10$ agents through $35$ agents, corresponding to $21$ and $71$ levels, respectively.
For example, in the $\N = 10$ configuration with no failures ($\NF = 0$), the THD of the array is around $5\%$.
In the $\N = 15$ configuration with $\NF = 5$ failures, the THD is also around $5\%$.
These configurations may result in too high a THD for the grid-tie, but the THD is around $2.5\%$ for $\NNF \geq 16$, so as long as there are at least a large fraction of functioning panels and inverter modules in large arrays, the grid-tie could be connected.

\begin{figure}[t!]
	\centering
	\includegraphics[width=\columnwidth]{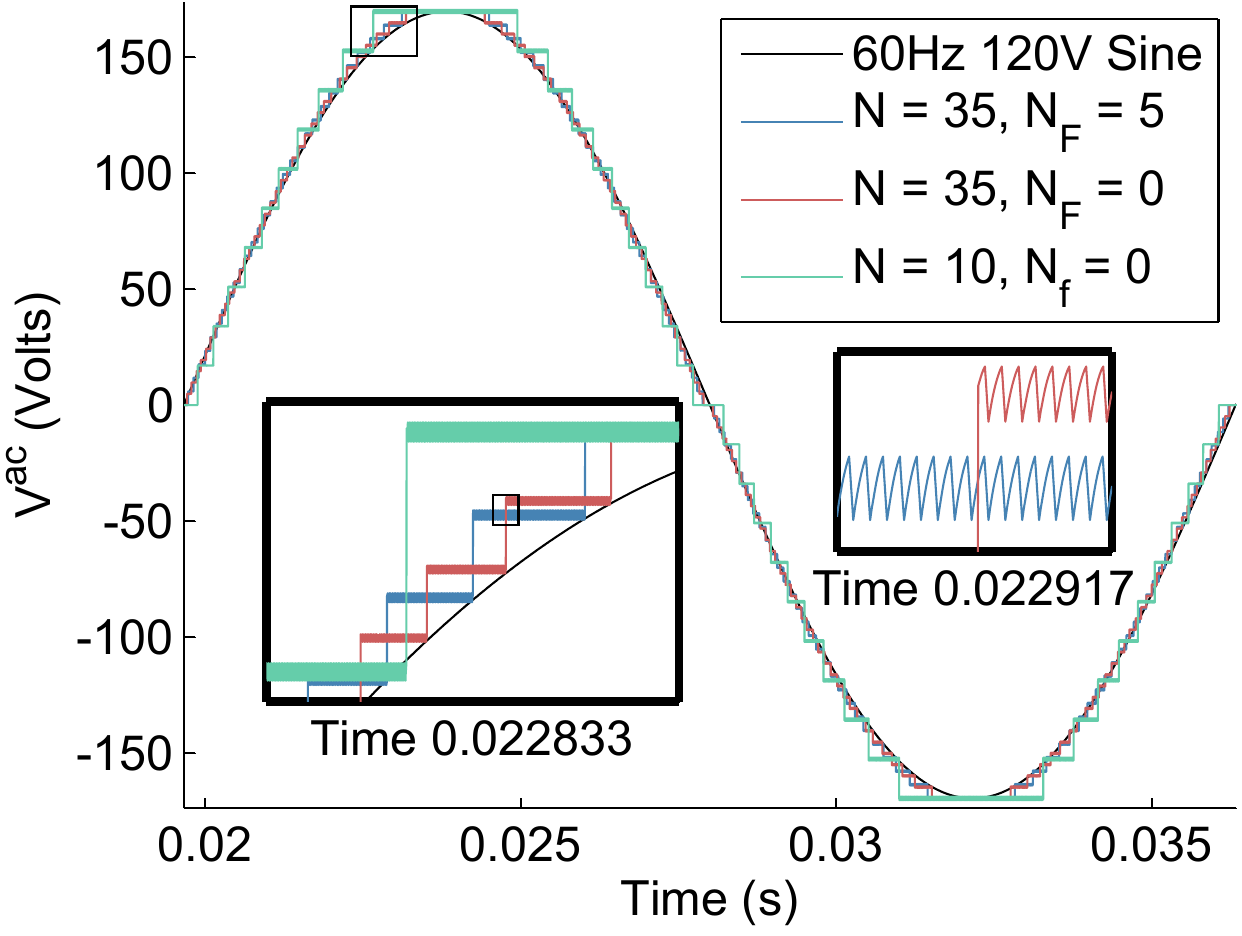}
	\vspace{-0.5em}
	\caption{Executions of three configurations of the array, with $\N = 35$ agents and $\NF = 5$ failures, with $\N = 35$ agents and no failures, and $\N = 10$ agents and no failures.  The figures illustrate the different H-bridge switching times and buck-boost regulated voltage levels in different configurations.}
	\figlabel{execution_vac}
	\vspace{-0.5em}
\end{figure}

\begin{figure}[t!]
	\centering
	\includegraphics[width=0.9\columnwidth]{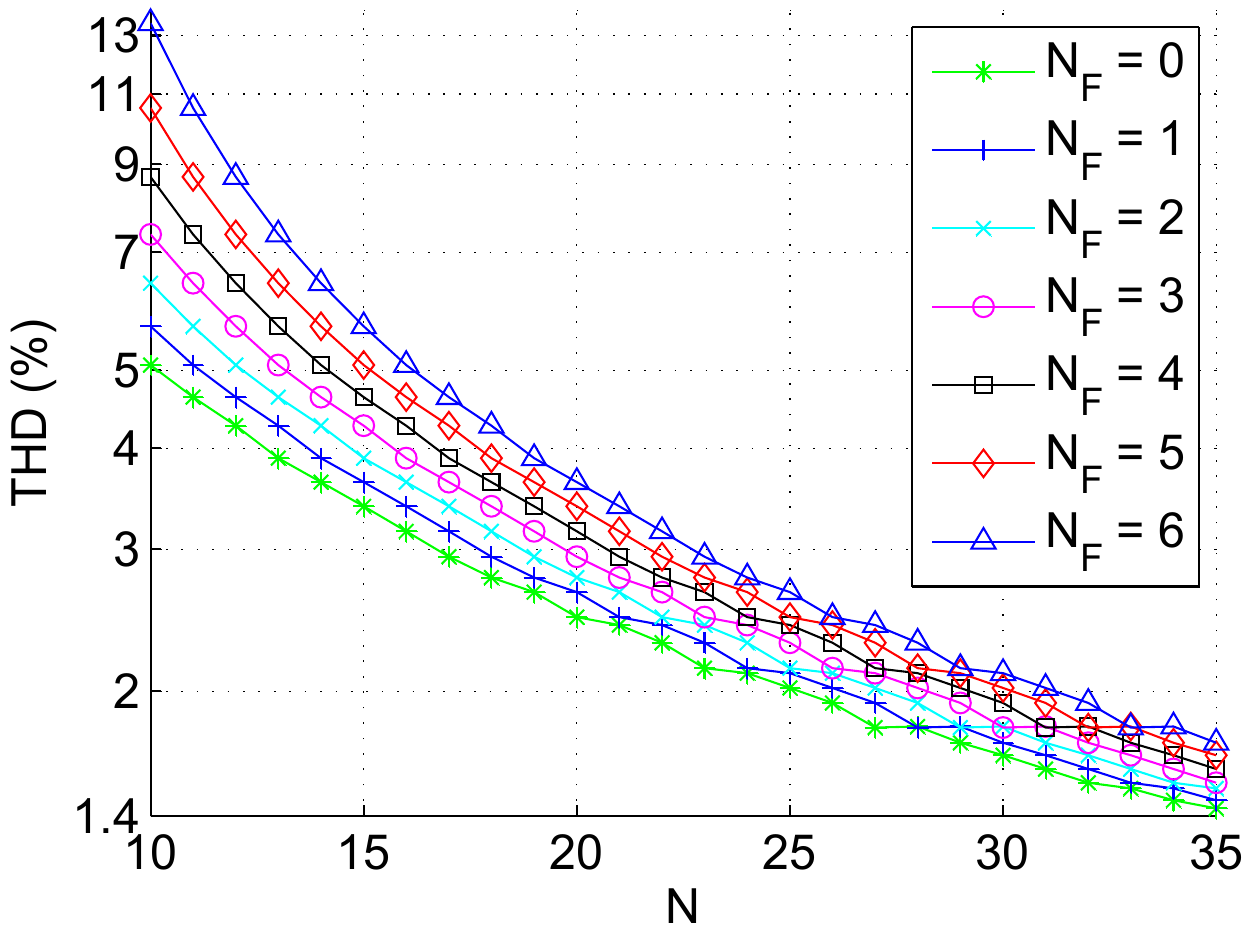}
	\vspace{-0.5em}
	\caption{THD for different array configurations consisting of $\N$ panels and inverter modules (agents), along with different numbers of statically failed agents $\NF$ at system start-up.  The $y$-axis scale is logarithmic.}
	\figlabel{thd_failure}
	\vspace{-1.5em}
\end{figure}

\subsection{THD with Dynamic Failures}
Dynamic failures are those that occur once the grid-tie is operational and connected.
We consider dynamic failures ($\NF = 1$) of one agent at a time.
\figreftwo{execution_vac_failure}{execution_vac_failure_N30} each, respectively, show the grid-tie voltage $\Vac$ versus time for three executions with one random dynamic failure that occurs at a uniformly distributed random time in the period.
These scenarios are considered as failures at different times result in varying performance degradation of the THD.
For instance, one scenario is where a failure of an agent that is not connected to $\Vac$ at a time instant.
One hypothesis is that such a failure may not negatively impact the THD, as it is not connected to the output.
However, each of the remaining operational agents $i \in \IDNF$ must (a) increase their output voltages $\Vdci$ since there is one fewer level, and (b) change their H-bridge switching times $\twaitiset$ using the algorithm of~\eqref{hbridge_switch_local}.
\figref{execution_vdc_failure} shows the H-bridge output voltage $\Vaci$ for each agent $i \in \ID$ for a configuration with $\N = 6$ agents and one dynamic failure.

\begin{figure}[t!]
	\centering
	\includegraphics[width=0.95\columnwidth]{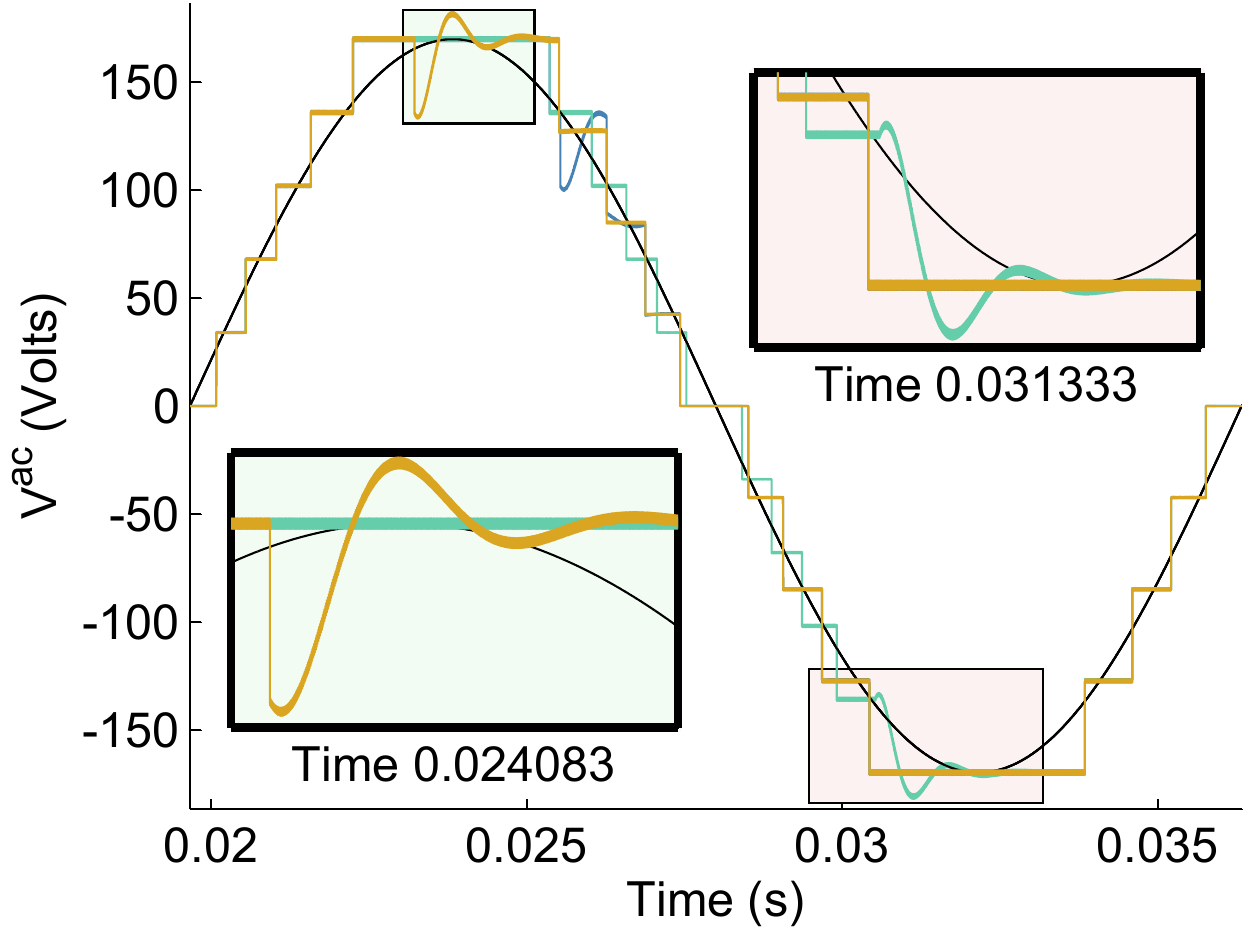}
	\vspace{-0.5em}
	\caption{The black line is an ideal $60$ Hz $120$ V sine, and the green, yellow, and blue lines are each an execution of $\A$ with $\N = 5$ agents and $1$ dynamic failure at a different random time.  The failure causes the total number of voltage levels to transition from $(2\N+1) = 11$ to $(2\NNF+1) = 9$ levels.  The zoom plots illustrate the fast recovery as the buck-boost converter reference voltage control and the H-bridges' switching times are changed.  Note that in each of the three executions, in the first quarter-period ($t \leq 0.022$) there are $\N = 5$ positive voltage levels as there are $5$ functioning agents, and the recovery is fast enough that by fourth quarter-period ($t \geq 0.0325$) there are $\N = 4$ negative voltage levels due to the one dynamic failure ($\NF = 1$).}
	\figlabel{execution_vac_failure}
	\vspace{-1em}
\end{figure}

\begin{figure}[t!]
	\vspace{-0.5em}
	\centering
	\includegraphics[width=0.95\columnwidth]{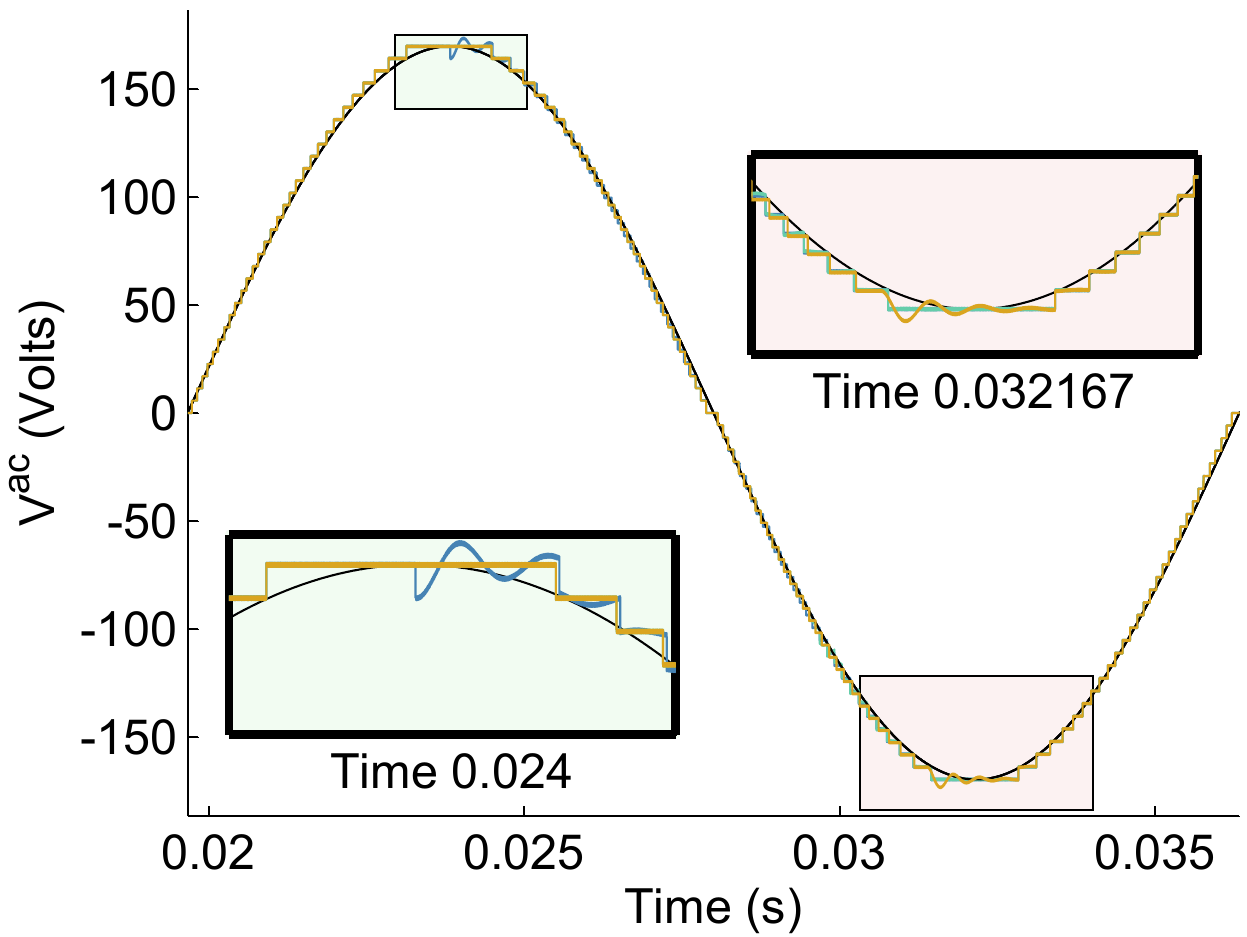}
	\vspace{-0.5em}
	\caption{The black line is an ideal $60$ Hz $120$ V sine, and the green, yellow, and blue lines are each an execution of $\A$ with $\N = 30$ agents and $1$ dynamic failure at a different random time.  The failure causes the total number of voltage levels to transition from $(2\N+1) = 61$ to $(2\NNF+1) = 59$ levels.  The zoom plots illustrate the fast recovery as the buck-boost converter reference voltage control and the H-bridges' switching times are changed.  Note that in each of the three executions, in the first quarter-period ($t \leq 0.022$) there are $\N = 30$ positive voltage levels as there are $30$ functioning agents, and the recovery is fast enough that by fourth quarter-period ($t \geq 0.0325$) there are $\N = 29$ negative voltage levels due to the one dynamic failure ($\NF = 1$).}
	\figlabel{execution_vac_failure_N30}
	\vspace{-1em}
\end{figure}

\begin{figure}[t!]
	\centering
	\includegraphics[width=0.95\columnwidth]{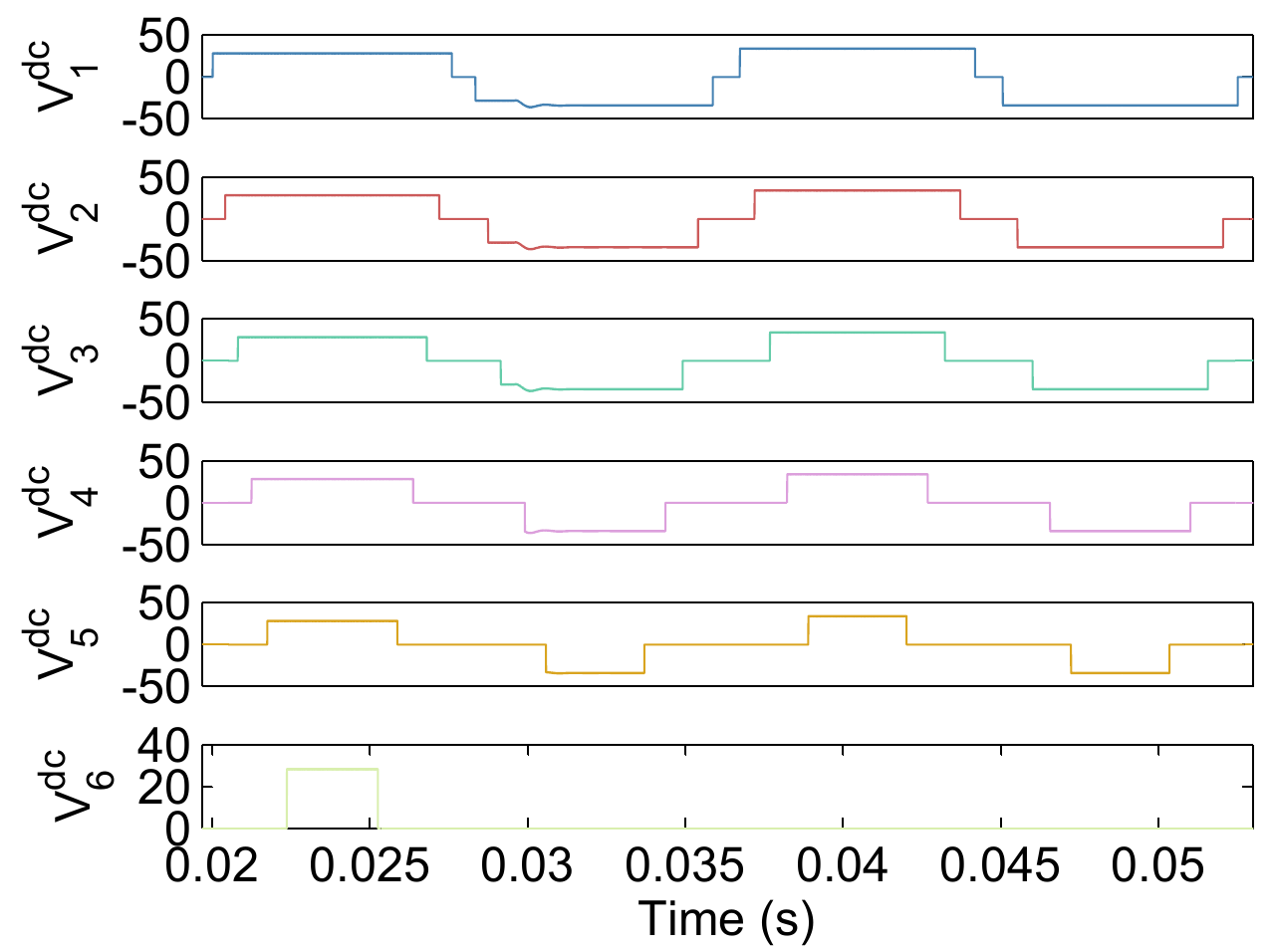}
	\vspace{-0.5em}
	\caption{Execution of $\N = 6$ agents with one dynamic failure ($\NF = 1$) illustrating the H-bridge switching signals and output voltages $\Vdci$ for each panel $i \in \ID$.  The failure causes panel $i = 6$ to have $\failedi = \true$ and $\Vdci = 0$, so the operating agents $i \in \IDNF$ increase their reference voltages $\Vrefi$ and update their switching times $\twaitiset$.}
	\figlabel{execution_vdc_failure}
	\vspace{-1em}
\end{figure}

\figref{execution_thd_failure_dynamic} shows averaged THD versus time over two periods ($2\Tgrid$) for arrays composed of $\N = 5$ to $35$ agents in increments of $5$ agents where a single dynamic failure ($\NF = 1$) occurs in the first of the two periods.
These results correspond to the scenarios depicted in~\figreftwo{execution_vac_failure}{execution_vac_failure_N30}, with the averaged THD in~\figref{thd_failure_dynamic}.
\figref{execution_thd_failure_dynamic} indicates that in the case of a single failure, the THD of the $\N$ agent system returns to that of the $\N-1$ agent system quickly (within one period $\Tgrid$).
\begin{figure}[t!]
	\centering
	\includegraphics[width=0.9\columnwidth]{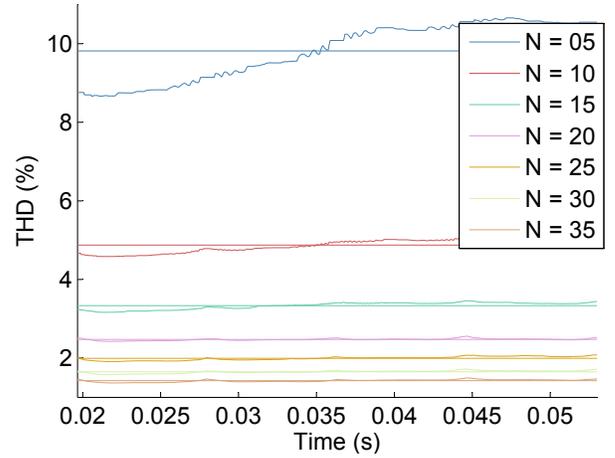}
	\vspace{-0.5em}
	\caption{Averaged THD versus time for the $20$ executions with uniformly sampled random failure time for $\A$ with $\N = 5$ through $\N = 35$ agents and $1$ dynamic failure at different random times.  The straight lines are the THD averages without failures (e.g., from~\figref{thd_failure} with $\NF = 0$), and the time-varying lines are the THD at different instances due to the dynamic failure.  The $y$-axis scale is logarithmic.}
	\vspace{-1em}
	\figlabel{execution_thd_failure_dynamic}
\end{figure}
It is unlikely more than a single dynamic failure would occur simultaneously before recovery, which as shown in~\figreffour{execution_vac_failure}{execution_vac_failure_N30}{execution_vdc_failure}{execution_thd_failure_dynamic}, happens in under half a grid period $\Tgrid$.
Furthermore, if one failure occurs, from our previous analysis of THD under static failures (\figref{thd_failure}), we see that the array behavior simply returns to the system's behavior with $\N - 1$ operating agents.
Thus, if more than a single dynamic failure occurs ($\NF > 1$, as long as each failure is spaced out enough in time (greater than a half grid period apart), the overall behavior will just return the array to the behavior with $\N - 1$, then $\N - 2$, $\ldots$, $\N - \NF$ panels operating.

\begin{figure}[t!]
	\centering
	\includegraphics[width=0.9\columnwidth]{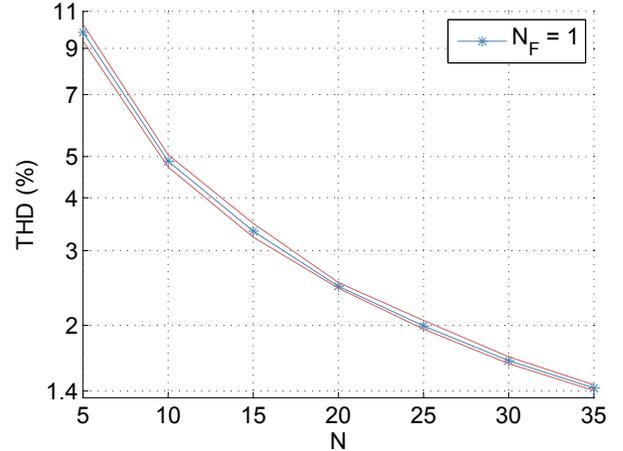}
	\vspace{-0.5em}
	\caption{Averaged THD of $20$ simulations for different array configurations consisting of $\N$ panels and inverter modules (agents) in increments of $5$ between $5$ and $35$, along with a one dynamically failed agent, $\NF = 1$, that fails at a uniformly-sampled random time in the grid period $\Tgrid$.  The $y$-axis scale is logarithmic, the center line is the mean, and the upper and lower lines are $\alpha = 95$ confidence intervals.}
	\vspace{-2em}
	\figlabel{thd_failure_dynamic}
\end{figure}

\section{Conclusion and Future Work}
\seclabel{conclusion}
In this paper, we have presented a model-based design and virtual prototyping analysis of a distributed inverter used as a grid-tie to connect $\N$ DC voltage sources, in this case solar panels, to the grid.
In addition to the solar array considered here, the design, failure modeling, and analysis may be useful in numerous scenarios using multilevel inverters as grid-ties.
In particular, the paper illustrates the feasibility of individual and multiple inverter modules failing in certain ways, and being able to keep the grid-tie operational with acceptable performance deterioration (in terms of THD).
In future work, we plan to construct an actual prototype of the array and evaluate its fault-tolerance capabilities in real-world scenarios.
For the actual prototype, we plan to employ a switching scheme to vary the switching times used by each agent's H-bridge to decrease wear by periodically changing identifiers of all the agents using a distributed identifier algorithm.
Additionally, we plan to formally verify several specifications of the H-bridge control algorithm regardless of the number of inverter modules, $\N$, using the \toolpassel verification tool~\cite{johnson2013phdthesis}.
For example, one basic specification is that the switching logic of the modules never results in modules with opposite polarity voltages being connected together for the grid tie.
This can be formulated as a verification problem for timed automata as done previously for an array with fixed size ($\N$) in~\cite{johnson2012peci}.



\scriptsize
\let\oldbibliography\thebibliography
\renewcommand{\thebibliography}[1]{\oldbibliography{#1}
\setlength{\itemsep}{0pt}} 
\bibliographystyle{IEEEtran}
\bibliography{IEEEabrv,master,luan}  

\begin{thebibliography}{10}
\providecommand{\url}[1]{#1}
\csname url@samestyle\endcsname
\providecommand{\newblock}{\relax}
\providecommand{\bibinfo}[2]{#2}
\providecommand{\BIBentrySTDinterwordspacing}{\spaceskip=0pt\relax}
\providecommand{\BIBentryALTinterwordstretchfactor}{4}
\providecommand{\BIBentryALTinterwordspacing}{\spaceskip=\fontdimen2\font plus
\BIBentryALTinterwordstretchfactor\fontdimen3\font minus
  \fontdimen4\font\relax}
\providecommand{\BIBforeignlanguage}[2]{{%
\expandafter\ifx\csname l@#1\endcsname\relax
\typeout{** WARNING: IEEEtran.bst: No hyphenation pattern has been}%
\typeout{** loaded for the language `#1'. Using the pattern for}%
\typeout{** the default language instead.}%
\else
\language=\csname l@#1\endcsname
\fi
#2}}
\providecommand{\BIBdecl}{\relax}
\BIBdecl

\bibitem{lai1996}
J.-S. Lai and F.~Z. Peng, ``Multilevel converters-a new breed of power
  converters,'' \emph{Industry Applications, IEEE Transactions on}, vol.~32,
  no.~3, pp. 509--517, May 1996.

\bibitem{tolbert1999}
L.~Tolbert, F.~Z. Peng, and T.~Habetler, ``Multilevel converters for large
  electric drives,'' \emph{Industry Applications, IEEE Transactions on},
  vol.~35, no.~1, pp. 36--44, Jan. 1999.

\bibitem{mcgrath2002tie}
B.~McGrath and D.~Holmes, ``Multicarrier {PWM} strategies for multilevel
  inverters,'' \emph{Industrial Electronics, IEEE Transactions on}, vol.~49,
  no.~4, pp. 858--867, Aug. 2002.

\bibitem{kjaer2005tia}
S.~Kjaer, J.~Pedersen, and F.~Blaabjerg, ``A review of single-phase
  grid-connected inverters for photovoltaic modules,'' \emph{Industry
  Applications, IEEE Transactions on}, vol.~41, no.~5, pp. 1292--1306, 2005.

\bibitem{franquelo2008iem}
L.~Franquelo, J.~Rodriguez, J.~Leon, S.~Kouro, R.~Portillo, and M.~Prats, ``The
  age of multilevel converters arrives,'' \emph{Industrial Electronics
  Magazine, IEEE}, vol.~2, no.~2, pp. 28--39, June 2008.

\bibitem{cecati2010tii}
C.~Cecati, F.~Ciancetta, and P.~Siano, ``A multilevel inverter for photovoltaic
  systems with fuzzy logic control,'' \emph{Industrial Electronics, IEEE
  Transactions on}, vol.~57, no.~12, pp. 4115--4125, 2010.

\bibitem{johnson_brian2013phdthesis}
B.~Johnson, ``Control, analysis, and design of distributed inverter systems,''
  Ph.D. dissertation, University of Illinois at Urbana-Champaign, 2013.

\bibitem{filho2011tia}
F.~Filho, L.~Tolbert, Y.~Cao, and B.~Ozpineci, ``Real-time selective harmonic
  minimization for multilevel inverters connected to solar panels using
  artificial neural network angle generation,'' \emph{Industry Applications,
  IEEE Transactions on}, vol.~47, no.~5, pp. 2117--2124, 2011.

\bibitem{turpin2002tie}
C.~Turpin, P.~Baudesson, F.~Richardeau, F.~Forest, and T.~Meynard, ``Fault
  management of multicell converters,'' \emph{Industrial Electronics, IEEE
  Transactions on}, vol.~49, no.~5, pp. 988--997, Oct. 2002.

\bibitem{chen2005tpe}
A.~Chen, L.~Hu, L.~Chen, Y.~Deng, and X.~He, ``A multilevel converter topology
  with fault-tolerant ability,'' \emph{Power Electronics, IEEE Transactions
  on}, vol.~20, no.~2, pp. 405--415, 2005.

\bibitem{chan2006ciep}
F.~Chan and H.~Calleja, ``Reliability: A new approach in design of inverters
  for pv systems,'' in \emph{International Power Electronics Congress, 10th
  IEEE}, Oct. 2006, pp. 1--6.

\bibitem{ma2007tpe}
M.~Ma, L.~Hu, A.~Chen, and X.~He, ``Reconfiguration of carrier-based modulation
  strategy for fault tolerant multilevel inverters,'' \emph{Power Electronics,
  IEEE Transactions on}, vol.~22, no.~5, pp. 2050--2060, Sep. 2007.

\bibitem{khomfoi2007tpe}
S.~Khomfoi and L.~Tolbert, ``Fault diagnostic system for a multilevel inverter
  using a neural network,'' \emph{Power Electronics, IEEE Transactions on},
  vol.~22, no.~3, pp. 1062--1069, May 2007.

\bibitem{ristow2008tie}
A.~Ristow, M.~Begovic, A.~Pregelj, and A.~Rohatgi, ``Development of a
  methodology for improving photovoltaic inverter reliability,''
  \emph{Industrial Electronics, IEEE Transactions on}, vol.~55, no.~7, pp.
  2581--2592, Jul. 2008.

\bibitem{lezana2010tie}
P.~Lezana, J.~Pou, T.~Meynard, J.~Rodriguez, S.~Ceballos, and F.~Richardeau,
  ``Survey on fault operation on multilevel inverters,'' \emph{Industrial
  Electronics, IEEE Transactions on}, vol.~57, no.~7, pp. 2207--2218, Jul.
  2010.

\bibitem{song2013tpe}
Y.~Song and B.~Wang, ``Survey on reliability of power electronic systems,''
  \emph{Power Electronics, IEEE Transactions on}, vol.~28, no.~1, pp. 591--604,
  Jan. 2013.

\bibitem{harb2013tpe}
S.~Harb and R.~Balog, ``Reliability of candidate photovoltaic
  module-integrated-inverter (pv-mii) topologies--a usage model approach,''
  \emph{Power Electronics, IEEE Transactions on}, vol.~28, no.~6, pp.
  3019--3027, 2013.

\bibitem{zhao2013tii}
Y.~Zhao, J.~de~Palma, J.~Mosesian, R.~Lyons, and B.~Lehman, ``Line-line fault
  analysis and protection challenges in solar photovoltaic arrays,''
  \emph{Industrial Electronics, IEEE Transactions on}, vol.~60, no.~9, pp.
  3784--3795, 2013.

\bibitem{fischer2014tii}
J.~Fischer, S.~Gonzalez, M.~Herran, M.~Judewicz, and D.~Carrica,
  ``Calculation-delay tolerant predictive current controller for three-phase
  inverters,'' \emph{Industrial Informatics, IEEE Transactions on}, vol.~10,
  no.~1, pp. 233--242, 2014.

\bibitem{dhople2012tpe}
S.~Dhople, A.~Davoudi, A.~Dominguez-Garcia, and P.~Chapman, ``A unified
  approach to reliability assessment of multiphase dc-dc converters in
  photovoltaic energy conversion systems,'' \emph{Power Electronics, IEEE
  Transactions on}, vol.~27, no.~2, pp. 739--751, Feb 2012.

\bibitem{bolognani2000experimental}
S.~Bolognani, M.~Zordan, and M.~Zigliotto, ``Experimental fault-tolerant
  control of a pmsm drive,'' \emph{Industrial Electronics, IEEE Transactions
  on}, vol.~47, no.~5, pp. 1134--1141, 2000.

\bibitem{kou2002full}
X.~Kou, K.~A. Corzine, and Y.~L. Familiant, ``Full binary combination schema
  for floating voltage source multi-level inverters,'' in \emph{Industry
  Applications Conference, 2002. 37th IAS Annual Meeting. Conference Record of
  the}, vol.~4.\hskip 1em plus 0.5em minus 0.4em\relax IEEE, 2002, pp.
  2398--2404.

\bibitem{johnson2011apec}
B.~Johnson, P.~Krein, and P.~Chapman, ``Photovoltaic ac module composed of a
  very large number of interleaved inverters,'' in \emph{Applied Power
  Electronics Conference and Exposition (APEC), 2011 Twenty-Sixth Annual IEEE},
  Mar. 2011, pp. 976--981.

\bibitem{esram2007tec}
T.~Esram and P.~Chapman, ``Comparison of photovoltaic array maximum power point
  tracking techniques,'' \emph{Energy Conversion, IEEE Transactions on},
  vol.~22, no.~2, pp. 439--449, June 2007.

\bibitem{patel2008tec}
H.~Patel and V.~Agarwal, ``{MATLAB}-based modeling to study the effects of
  partial shading on {PV} array characteristics,'' \emph{Energy Conversion,
  IEEE Transactions on}, vol.~23, no.~1, pp. 302--310, Mar. 2008.

\bibitem{ropp2009tec}
M.~Ropp and S.~Gonzalez, ``Development of a {MATLAB}/{S}imulink model of a
  single-phase grid-connected photovoltaic system,'' \emph{Energy Conversion,
  IEEE Transactions on}, vol.~24, no.~1, pp. 195--202, Mar. 2009.

\bibitem{ding2012tec}
K.~Ding, X.~Bian, H.~Liu, and T.~Peng, ``A {MATLAB}-{S}imulink-based {PV}
  module model and its application under conditions of nonuniform irradiance,''
  \emph{Energy Conversion, IEEE Transactions on}, vol.~27, no.~4, pp. 864--872,
  Dec. 2012.

\bibitem{maki2013tec}
A.~Maki and S.~Valkealahti, ``Effect of photovoltaic generator components on
  the number of {MPP}s under partial shading conditions,'' \emph{Energy
  Conversion, IEEE Transactions on}, vol.~28, no.~4, pp. 1008--1017, Dec. 2013.

\bibitem{alur1995tcs}
R.~Alur, C.~Courcoubetis, N.~Halbwachs, T.~A. Henzinger, P.-H. Ho, X.~Nicollin,
  A.~Olivero, J.~Sifakis, and S.~Yovine, ``The algorithmic analysis of hybrid
  systems,'' \emph{Theoretical Computer Science}, vol. 138, no.~1, pp. 3--34,
  1995.

\bibitem{lynch2003ic}
N.~Lynch, R.~Segala, and F.~Vaandrager, ``Hybrid {I/O} automata,''
  \emph{Information and Computation}, vol. 185, no.~1, pp. 105--157, 2003.

\bibitem{kaynar2006book}
D.~K. Kaynar, N.~Lynch, R.~Segala, and F.~Vaandrager, \emph{The Theory of Timed
  {I/O} Automata}, ser. Synthesis Lectures in Computer Science.\hskip 1em plus
  0.5em minus 0.4em\relax Morgan \& Claypool, 2006.

\bibitem{johnson2013phdthesis}
T.~T. Johnson, ``Uniform verification of safety for parameterized networks of
  hybrid automata,'' Ph.D. dissertation, University of Illinois at
  Urbana-Champaign, Urbana, IL 61801, 2013.

\bibitem{johnson2012peci}
T.~T. Johnson, Z.~Hong, and A.~Kapoor, ``Design verification methods for
  switching power converters,'' in \emph{Power and Energy Conference at
  Illinois (PECI), 2012 IEEE}, Feb. 2012, pp. 1--6.

\bibitem{hossain2013peci}
S.~Hossain, S.~Dhople, and T.~T. Johnson, ``Reachability analysis of
  closed-loop switching power converters,'' in \emph{Power and Energy
  Conference at Illinois (PECI)}, 2013, pp. 130--134.

\bibitem{nguyen2014arch}
L.~V. Nguyen and T.~T. Johnson, ``Benchmark: Dc-to-dc switched-mode power
  converters (buck converters, boost converters, and buck-boost converters),''
  in \emph{Applied Verification for Continuous and Hybrid Systems Workshop
  (ARCH 2014)}, Berlin, Germany, Apr. 2014.

\bibitem{althoff2014tps}
M.~Althoff, ``Formal and compositional analysis of power systems using
  reachable sets,'' \emph{Power Systems, IEEE Transactions on}, no.~99, pp.
  1--11, 2014.

\bibitem{johnson2010icdcs}
T.~T. Johnson, S.~Mitra, and K.~Manamcheri, ``Safe and stabilizing distributed
  cellular flows,'' in \emph{Proceedings of the 30th {IEEE} International
  Conference on Distributed Computing Systems (ICDCS)}.\hskip 1em plus 0.5em
  minus 0.4em\relax Genoa, Italy: IEEE, June 2010.

\bibitem{johnson2011jnsa}
T.~T. Johnson and S.~Mitra, ``Safe flocking in spite of actuator faults using
  directional failure detectors,'' \emph{Journal of Nonlinear Systems and
  Applications}, vol.~2, no. 1-2, pp. 73--95, Apr. 2011.

\bibitem{lynch1996book}
N.~A. Lynch, \emph{Distributed Algorithms}.\hskip 1em plus 0.5em minus
  0.4em\relax San Francisco, CA, USA: Morgan Kaufmann Publishers Inc., 1996.

\bibitem{severns1985book}
R.~P. Severns and G.~Bloom, \emph{Modern DC-to-DC Switchmode Power Converter
  Circuits}.\hskip 1em plus 0.5em minus 0.4em\relax New York, New York: Van
  Nostrand Reinhold Company, 1985.

\bibitem{erickson2004book}
R.~W. Erickson and D.~Maksimovi\'{c}, \emph{Fundamentals of Power Electronics},
  2nd~ed.\hskip 1em plus 0.5em minus 0.4em\relax Springer, 2004.

\end{thebibliography}


\begin{IEEEbiography}{\authorluan}
Biography to be added in a final version if accepted.
\end{IEEEbiography}

\begin{IEEEbiography}{\authortaylor}
Biography to be added in a final version if accepted.
\end{IEEEbiography}
%






\end{document}